\documentclass[aps,prb,groupedaddress,superscriptaddress,reprint,twocolumn]{revtex4-2}
\usepackage{hyperref}
\hypersetup{colorlinks=true, citecolor=blue, urlcolor=blue, linkcolor=blue}
\usepackage{amsfonts}
\usepackage{amsmath}
\usepackage{amssymb}
\usepackage{graphicx}%
\usepackage{dcolumn}

\begin{document}


\title{Parameter-free quantum hydrodynamic theory for plasmonics: Electron density-dependent damping rate and diffusion coefficient}


\author{Qi-Hong Hu}
\affiliation{Department of Physics, Jishou University, Jishou 416000, People’s Republic of China}
\author{Ren-Feng Liu}
\affiliation{Department of Physics, Jishou University, Jishou 416000, People’s Republic of China}
\author{Xin-Yu Shan}
\affiliation{Department of Physics, Jishou University, Jishou 416000, People’s Republic of China}
\author{Xuan-Ren Chen}
\affiliation{Department of Physics, Jishou University, Jishou 416000, People’s Republic of China}
\author{Hong Yang}
\email{yanghong@jsu.edu.cn}
\affiliation{Department of Physics, Jishou University, Jishou 416000, People’s Republic of China}
\author{Peng Kong}
\affiliation{Department of Physics, Jishou University, Jishou 416000, People’s Republic of China}
\author{Xiao-Yun Wang}
\affiliation{Department of Physics, Jishou University, Jishou 416000, People’s Republic of China}
\author{Ke Deng}
\affiliation{Department of Physics, Jishou University, Jishou 416000, People’s Republic of China}
\author{Xiangyang Peng}
\affiliation{Hunan Key Laboratory for Micro-Nano Energy Materials and Devices, School of Physics and Optoelectronics, Xiangtan University, Hunan 411105, People’s Republic of China}
\author{Dong Xiang}
\affiliation{Institute of Modern Optics and Center of Single Molecule Sciences, Nankai University, Key Laboratory of Micro-scale Optical Information Science and Technology, Tianjin 300350, China} 
\author{Yong-Gang Huang}
\email{huang122012@163.com}
\affiliation{Department of Physics, Jishou University, Jishou 416000, People’s Republic of China}

\date{\today}

\begin{abstract}
Plasmonics is a rapid growing field, which has enabled both exciting, new fundamental science and inventions of various quantum optoelectronic devices. An accurate and efficient method to calculate the optical response of metallic structures with feature size in the nanoscale plays an important role in plasmonics. Quantum hydrodynamic theory (QHT) provides an efficient description of the free-electron gas, where quantum effects of nonlocality and spill-out are taken into account. In this work, we introduce a general QHT that includes diffusion to account for the size-dependent broadening, which is a key problem in practical applications of surface plasmon. We will introduce a density-dependent diffusion coefficient to give very accurate linewidth. It is a self-consistent method, in which both the ground and excited states are solved by using the same energy functional, with the kinetic energy described by the Thomas-Fermi (TF) and von Weizs\"{a}cker (vW) formalisms. We numerically prove that the fraction of the vW should be around $0.4$. In addition, our QHT method is stable by introduction of an electron density-dependent damping rate. For sodium nanosphere of various sizes, the plasmon energy and broadening by our QHT method are in excellent agreement with those by density functional theory and Kreibig formula. By applying our QHT method to sodium jellium nanorods of various sizes, we clearly show that our method enables a parameter-free simulation, i.e. without resorting to any empirical parameter such as size-dependent damping rate, diffusing coefficient and the fraction of the vW. It is found that there exists a perfect linear relation between the main longitudinal localized surface plasmon resonance wavelength and the aspect radio. The width decreases with increasing aspect ratio and height. The calculations show that our QHT method provides an explicit and unified way to account for size-dependent frequency shifts and broadening of arbitrarily shaped geometries. It is reliable and robust with great predicability, and hence provides a general and efficient platform to study plasmonics.  
\end{abstract}


\maketitle

\section{INTRODUCTION}
Plasmonic nanostructure can be used to reduce the size of optical devices and enhance the light-matter interaction for their ability to localize electromagnetic field well below the diffraction limit \cite{Schuller2010Plasmonics,Gramotnev2010Plasmonics,Bartal2008A,Baranov2017Novel,Kinkhabwala2009Large,PhysRevLett.118.237401,PhysRevLett.118.073604,Ringler2008Shaping,PhysRevLett.102.146807,Baranov2017Novel,https://doi.org/10.1002/lpor.201300144,PhysRevLett.97.053002,2016Single,akimov2007generation}. Many novel phenomena have been reported, such as quantum emitter-plasmon bound state \cite{Wen:20,Karanikolas:21}, reversible decay dynamics \cite{PhysRevA.99.053844,doi:10.1063/5.0033531}, polarization dependence of fluorescence \cite{doi:10.1021/nl902095q,Curto930}, position dependent dipole-dipole interaction \cite{dan2015position}, enhanced solar energy conversion \cite{Suljo2011Plasmonic,Cushing2016Progress}, biomedicine \cite{Min2006Gold,Huang2008Plasmonic,doi:10.1063/5.0065833}, surface-enhanced Raman scattering \cite{Yang_2021,https://doi.org/10.1002/anie.201205748}, plasmon rulers with ultrahigh sensitivity \cite{Wen2018Probing}, plasmonic photocatalysis \cite{Zhang2013Plasmonic}, plasmonic nanoantennas \cite{doi:10.1021/cr1002672}, sensors \cite{C9NR08433A}, plasmon laser \cite{Oulton2009Plasmon}, nano-optical tweezers \cite{2011Plasmon}, etc. 

Theoretically, the classical Drude model under the local response approximation (LRA) for the free electron is usually applied to understand the optical response of plasmonic nanostructure. However, when the characteristic size of the nanostructure falls below $10\,nm$ \cite{2012Quantum,ZHOU20191,CiracJurgaKhalidDellaSala+2019+1821+1833} or when the gap size of nanodimers becomes subnanometre \cite{2016Quantum,2012Bridging,doi:10.1021/acs.jpcc.7b07462}, the LRA breaks down due to its neglect of quantum effects such as nonlocality, electronic spill-out, and Landau damping. In principle, time-dependent density functional theory (TD-DFT) \cite{ullrich2011time,doi:10.1021/cr100265f} may be used to describe plasmon excitations in a quantum mechanical setting. However, a full quantum treatment of optical response is possible only for very small  cluster of a few atoms or highly symmetric nanostructure smaller than $2\,nm$ with pseudopotentials \cite{2006Optical}. Based on the jellium model without considering the atomic structure, up to 5000 electrons in nanosphere can be treated \cite{2019Plasmonic}. Although the problem is greatly simplified to one dimension by applying the spherical symmetry, the computation is still extremely costly. 

An alternative approach to the theoretical description of the free electron is the hydrodynamic theory. The collective motion of electrons in an arbitrary inhomogeneous system is expressed in terms of the average densities of electron and electron current \cite{Pitarke_2006,PhysRevB.84.121412,PhysRevB.95.245434}. Based on the perturbation theory, the linearized equation for the electron current density can be derived from the functional derivatives of the internal energy of the electron gas. When the internal kinetic energy (KE) is described by the Thomas-Fermi model and the electrons are strictly confined in the metallic structure, it is termed as the Thomas-Fermi hydrodynamic theory (TF-HT). In this model, the quantum pressure-related nonlocal response is taken into account, which is helpful to explain the blueshift of the surface plasmon resonance of silver nanoparticles with the size decreasing \cite{RazaStengerKadkhodazadehFischerKosteshaJauhoBurrowsWubsMortensen+2013+131+138}. Similar to the Drude model, the line broadening in TF-HT is determined through a phenomenological damping rate. For nanosphere with radius $R$, a so-called Kreibig term can be added to the bulk term \cite{1969The} in order to account for the size-dependent broadening, i.e. $\gamma=\gamma_0+Av_F/R$. However, it can not describe the line broadening for higher-order modes \cite{Raza_2015}. Besides, for complex-shaped nanostructure lacking spherical symmetry, it is hard to give such a Kreibig term to describe the size-dependent broadening. One treatment of these problems is the inclusion of diffusion currents of the conduction electrons in the TF-HT, which is termed as the generalized non-local optical response (GNOR) model \cite{Mortensen2014A}. As is stated in Ref. \cite{Mortensen+2021+2563+2616,Svendsen_2020}, this diffusion captures the effects of both mutual interactions among the electrons and the scattering of the electrons on metal surfaces, which mimics the surface-enhanced Landau damping due to the creation of electron–hole pairs. 

Although the convection-diffusion mechanism within the GNOR model provides a description of size-dependent frequency blueshift and linewidth broadening of the localized surface plasmon (LSP) resonance with decreasing particle size, it can not be directly applied to alkali metals where redshift was found. It has been found that the quantum mechanical effect of spill-out of electrons plays an important role. In addition, for two nanostructures with extremely narrow gap sizes, the electron spill-out leads to the overlap of electron in the gap region, which is vital to understand the strongly gap-dependent resonance energies and electric-field enhancements. QHT includes this quantum mechanical effect of spill-out \cite{toscano2015resonance,PhysRevB.91.115416}. In this case, the static density of electrons, obtained from a previous Kohn-Sham (KS) density functional theory (DFT) or QHT, shows fast decay around the metal surface and in the electron-tail region \cite{PhysRevB.1.4555,PhysRevB.91.115416,toscano2015resonance}. Due to this inhomogeneity, gradient corrections to the energy functional may nonetheless become significant \cite{PhysRev.136.B864}. The von Weizs\"{a} cker (vW) term is the leading-order correction to the TF kinetic energy, which should be added to avoid a vanishing work function \cite{PhysRevB.91.115416}. 

However, the fraction of the vW contribution ($\lambda_w$) is not well defined and usually in the range from $1/9$ to $1$.  For example, the best choice for $\lambda_w$ should be $0.6$ in order to give a small mean absolute relative error for the cell volume, bulk modulus, total energy at equilibrium volume, and density error for infinite sodium \cite{doi:10.1021/acs.jpclett.8b01926}. For semi-infinite sodium, the larger $\lambda_w$ leads to the larger work function ($\propto \lambda_w$) and it should be around $0.435$ to give the work function close to the DFT value \cite{PhysRevB.91.115416}. In addition, it was suggested that $\lambda_w$ should depends on excitation frequency in the bulk region, i.e. with $\lambda_w=1/9$ and $\lambda_w=1$ corresponding to low and high excitation frequency, respectively. In the density tail region,  $\lambda_w\rightarrow 1$ should be used. For sodium nanosphere, numerical tests show that it can control the degree of the electron spill-out, with small $\lambda_w$ corresponding to a less spill-out \cite{Li:15}, which should affect much the resonance frequency ($\propto \sqrt{1-N_{out}/N_e}$ with $N_{out}/N_e$ being a fraction of the spill-out electrons) \cite{PhysRevB.74.165421}.  In Ref. \cite{toscano2015resonance},  $\lambda_w=1/9$ was used for sodium and silver nanospheres.

Recently, there has been significant activity by Cirac\`{\i} and co-workers in QHT \cite{PhysRevB.93.205405,PhysRevX.11.011049}. In Ref. \cite{PhysRevB.93.205405}, they have shown that the energy of main LSP resonance for a sodium nanosphere is in good agreement with TD-DFT. $\lambda_w=1$ was used in the excited state calculation. Errors of about $20\,meV$ or $10\,meV$ for the resonance energy have been obtained, when the ground density is calculated by KS-DFT or given by an analytical model input density. To remove the computation-size dependent spurious peaks at energies higher than the main LSP resonance, the Laplacian-level KE functional was introduced in the electron density-tail region \cite{PhysRevX.11.011049}. It is found that this approach gives very accurate plasmon energy, peak intensity, and Feibelman $d$-parameter, as well as a single numerically stable Bennett state. However, for this method,  either the KS ground density or a model input density is required. It is computation expansive or can not be obtained for complex nonspherical nanostructures. In addition, similar to the Drude model and the TF-HT model, the line broadening was ‘put in by hand’. This is different from the GNOR model, in which the size-dependent broadening for nanostructure of arbitrary shape can be treated by the inclusion of diffusion current.  We emphasis that the 
diffusion can also be included in QHT, although the ground density is inhomogeneous. By using a density-dependent diffusion coefficient, we will show that the size-dependent broadening can be treated properly.

It should be noted that the KE used in Ref. \cite{PhysRevX.11.011049} can be well described by the Thomas-Fermi and von Weizs\"{a}cker (TF-vW) formalisms inside and around the nanosphere except for a small second-order correction (the Pauli-Gaussian formalism). Laplacian-level KE applied in this region leads to too large resonance energy. However, it should be applied in the extremely low density region in order to obtain a convergent result, which means that the TF-vW alone can not describe the physics there properly. Actually, the local plasmon frequency $ \omega_p=\sqrt {e^2 n_0/{m_e}{\varepsilon _0}}$ in the low density region ($n_0\rightarrow 0$) may be much smaller than the excitation frequency $\omega$, leading to large Landau damping associated with electron-hole pair generation \cite{khurgin2017landau,PhysRevLett.85.2200}.

Inspired by this observation, we introduce the diffusion current into the conventional QHT. Thus, both the convection–diffusion and spill-out effects are taken into account, which enables the QHT to provide a unified way to describe the size-dependent resonance energy and line broadening. In addition, we adopt the self consistent scheme \cite{toscano2015resonance}, in which the ground electron density is also determined by QHT. This enables the QHT to treat electronic response in relatively large-size nanosturcture of arbitrary shape. The TF KE functional with a fraction of vW correction $\lambda_w$ (i.e. the TF$\lambda_w$vW functional) will be used. A density-dependent damping with a large value in the density-tail region will be used to simulate the large Landau damping there, which will be helpful to remove the additional resonances above the main LSP resonance. The diffusion coefficient $D$ and the fraction of vW contribution $\lambda_w$ will be determined in order to give the resonance energy and broadening for sodium nanosphere of various sizes. We will show that our QHT enables a parameter-free simulation, which can be directly used to investigate both the ground and excited states properties of a generic electronic system without resorting to any empirical parameter, such as the damping rate, diffusing coefficient, and  $\lambda_w$. We apply this method to the study of optical response of sodium nanorod, which is an example of nonspherical shape. We will show that both the resonance energy and broadening predicted by our QHT are robust. 

\section{THEORY}
In this section, we present the QHT and numerically determine all the essential quantities in order to have a parameter-free form.  We will show that our QHT can predict very accurate resonance energy and line broadening for sodium nanospheres of various sizes. Both the ground and excited state will be solved by QHT with TF$\lambda_w$vW functional. This self consistent scheme enables its application to a general jellium system of arbitrary shape. In subsection A, we first present the conventional QHT and provide numerical details of its implementation. By applying to a sodium nanosphere,  we present the problems of conventional QHT, i.e. convergence problem and line broadening. In subsection B, we introduce a density-dependent damping rate and numerically show how it can be used to solve the convergence problem. In subsection C, we introduce a density-dependent diffusion current to form our parameter-free QHT. In the following subsection, we numerically determine all the required quantities, i.e. the fraction of vW contribution $\lambda_w$ and the coefficient $A$ for the diffusion term, in order to perform a parameter-free simulation.

\subsection{Convergence and linewidth by conventional quantum hydrodynamic theory}
The conventional linearized QHT response is governed by the following equations in the frequency domain \cite{PhysRevB.93.205405,PhysRevX.11.011049,PhysRevB.91.115416,toscano2015resonance}: 
\begin{subequations}
	\label{eq:whole}
	\begin{equation}	
     \nabla  \times \nabla \times \textbf{E}_s- \frac{{{\omega ^2}}}{{{c^2}}}{{\bf{E}}_s} = {\omega ^2}{\mu _0}{\bf{P}},\label{subeq:1}
	\end{equation}
	\begin{equation}
	\frac{e{n_0}}{m_e} \nabla \left( \dfrac{\delta G\left[ n \right]}{\delta n}\right)_1+	\left( {{\omega^2} + i\gamma \omega } \right){\bf{P}} =  - {\varepsilon_0} \omega_p^2 ({{\bf{E}}_i} + {{\bf{E}}_s})
	.\label{subeq:2}
	\end{equation}
\end{subequations}
Here ${\bf{E}}_i$ (${\bf{E}}_s$) and $\bf{P}$ are the incident (scattered) electric field and the polarization vector, respectively. $ c $, $ \varepsilon_0 $,  and $ \mu _0 $  are the speed of light, the permittivity, and the permeability in vacuum, respectively. $ m_e $  and $ e $  are the electron mass and charge.  $\gamma$ represents the phenomenological damping rate, which is an empirical parameter to account for the line broadening within the conventional QHT. In this work, it will be extended to a density-dependent quantity in order to solve the convergence problem, while the line broadening will be resolved by introducing diffusion electron current. $ \omega_p=\sqrt {e^2 n_0/{m_e}{\varepsilon _0}}$  is the plasma frequency with $ n_0 $ being the ground state electron density.  

To avoid using the KS ground density or an analytical model input density $n_0$, we follow the method presented in Ref. \cite{PhysRevB.91.115416,toscano2015resonance} with $n_0$ obtained in a self-consistent way. An advantage of this method is that numerical calculation is feasible for large nanostructure of arbitrary shape. The equation reads
\begin{equation}\label{ground}
\nabla^2 \left( \frac{\delta G\left[ n \right]}{{\delta n}} \right)_{0}  +\dfrac{e^2}{\varepsilon_0}(n_0-n_+)=0,
\end{equation}
with $n_+=(4\pi r_s^3/3)^{-1}$ being the positive charge density for the uniform jellium background.  In
this work, all calculations focus on sodium with Wigner-Seitz radius $r_s=4a_0$.  $a_0$ is the Bohr radius. $G \left[ n \right] $ is the quantum functional energy, which plays a central role in QHT. In essence, the QHT and the more advanced DFT and TD-DFT differ in those terms \cite{PhysRevB.91.115416}. The potentials
$ \left({\delta G\left[ n \right]}/{\delta n}\right)_0 $ and $ \left({\delta G\left[ n \right]}/{\delta n}\right)_1 $ refer to the unperturbed equilibrium case and the small nonequilibrium terms due to excitation, namely, ${\delta G\left[ n \right]}/{\delta n}=\left({\delta G\left[ n \right]}/{\delta n}\right)_0+\left({\delta G\left[ n \right]}/{\delta n}\right)_1$. The first-order term can be obtained using a perturbation approach where the perturbed density is taken as $n=n_0+n_1$, with $ {n_1(r)}=\nabla\cdot \textbf{P} /e$  being a small perturbation. The energy functional can be written as $ G\left[ n \right]= {T_s}\left[ n \right] + {E_{XC}^{LDA}}\left[ n \right] $, where ${E_{XC}^{LDA}}\left[ n \right]$ and $T_s$ are the exchange-correlation (XC) energy functional within the local density approximation (LDA) and the noninteracting KE functional, respectively. 

In this work, we use the kinetic energy functional of the form ${T_s}\left[ n \right] = {T_s}^{TF}\left[ n \right] + {\lambda _w}{T_s}^W\left[ n \right]$ ($\mathrm{TF}\lambda_w \mathrm{vW}$). As stated in the introduction, $\lambda _w $ is an important coefficient and will be determined. The expressions for the above potentials can be found in Ref. \cite{PhysRevB.93.205405} and references therein. Explicitly, they are 
\begin{subequations}
	\label{functionalderivative}
	\begin{equation}	
	\frac{\delta {T_s}^{TF}}{\delta n}=(E_h a_0^2) \dfrac{5}{3} c_{TF} n^{2/3},\label{subeq:TF}
	\end{equation}	
	\begin{equation}
	\frac{\delta {T_s}^{W}}{\delta n}=(E_h a_0^2) (\dfrac{1}{8} \dfrac{\triangledown n \cdot \triangledown n}{n^2} -2\dfrac{\triangledown^2 n}{n}), \label{subeq:vW}
	\end{equation}	
	\begin{equation}
	\frac{\delta E_{XC}^{LDA}}{\delta n}=(E_h)(-a_0\dfrac{4}{3}c_xn^{1/3}+\mu_c[n]), \label{subeq:XC}
	\end{equation}
\end{subequations}
where $E_h=\hbar/(m_e a_0^2)$ is the Hartree energy, $c_{TF}=\dfrac{3}{10}(3\pi^2)^{2/3}$, and $c_x=\frac{3}{4}(3/\pi)^{1/3}$.  The correlation potential $\mu_c[n]$  from the Perdew-Zunger LDA parametrization is $\frac{\alpha+7\alpha\beta_1\sqrt{r_{1}}/6+4\alpha\beta_2 r_{1}/3}{(1+\beta_1\sqrt{r_{1}}+\beta_2 r_{1})^2}$ with $r_{1}a_0 =(3/4\pi n)^{1/3}$, $\alpha=-0.1423$, $\beta_1=1.0529$, and $\beta_2=0.3334$.

\begin{figure}[htbp]
	\centering
	\includegraphics[width=4cm]{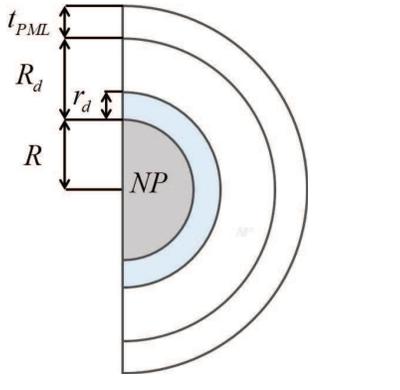}
	\caption{Schematic diagrams of the simulation domain. $R$ denotes the radius of the nanosphere, $r_d$ is the thickness for electron spill-out, ${R+R_d}$ denotes the simulation domain for $\mathbf{E}$, and $t_{pml}$ is the thickness of the perfectly matched layer (PML).}
	\label{fig1}%
\end{figure}

To demonstrate the performance of QHT, the normalized absorption cross section $\sigma/\sigma_0$ is calculated for a sodium nanosphere with radius $R$ excited by a plane wave. The absorption cross section can be calculated as $\sigma \left( \omega  \right)=\frac{\omega }{2{{I}_{0}}}\int{\left\{ \mathbf{E}\cdot {{\mathbf{P}}^{*}} \right\}dV}$ with ${{I}_{0}}\text{=}{{{\varepsilon }_{0}}cE_{i}^{2}}/{2}$ being the intensity for the incident plane wave $\mathbf{E}_i=\hat{z}E_ie^{ik_0x}$. $\mathbf{E}=\mathbf{E}_s+\mathbf{E}_i$ with $\mathbf{E}_s$ being the solution of Eq. \eqref{eq:whole}. ${{\sigma }_{0}}=\pi R^{2}$ is the geometrical area. The above differential equations [Eqs. \eqref{eq:whole} and \eqref{ground}] can be solved with a commercial software based on the finite-element method (FEM), COMSOL MULTIPHYSICS, which has been widely used in the plasmonic community, for example see Refs. \cite{Toscano:12,toscano2015resonance,PhysRevLett.107.096801,tian2019finite,yun2018renormalization,Zhao:18,PhysRevA.99.053844,Wen:20,PhysRevB.93.205405,PhysRevX.11.011049}. For axis symmetric structures, the 2.5D technique can be applied to reduce the computational cost \cite{Ciraci:13,PhysRevB.93.205405,PhysRevX.11.011049,tian2019finite}.  

Figure \ref{fig1} shows the schematic diagrams of the simulation domain for nanosphere system. $R=r_sN_e^{1/3}$ denotes the radius of the nanosphere.  $r_d$ is the electron spill-out thickness, which is an important parameter in the conventional QHT. $R+R_{d}$ denotes the radius of the simulation domain for the electric field, while $t_{pml}$ is the thickness of perfectly matched layer (PML) in order to emulate an infinite domain. In all the simulation without otherwise statement, the spill-out thickness for the excited state is $r_{d}=25a_0 $, while it is larger in the ground density calculation $r_{d}=50a_0$. We have checked that convergent results can be obtained by using $R_{d}=500a_0$, and $t_{PML}=200a_0$. The atomic units (a.u.) are used by setting $E_h=a_0=m_e=\hbar=1$ in all the expressions.  Forty mapped layers are used in the region $R-10a_0\leq r \leq R+10a_0$, i.e. in a shell with thickness $20a_0$ around the metal boundary, where the ground density varies greatly. Similarly, mapped layers with thickness $a_0$ are used in the region $R+10a_0\leq r \leq R+r_{d}$. A nonuniform mesh is employed with a maximum element size of $5a_0$ for the other metal area. For the rest of the computation domain, a mesh size of $40a_0$ suffices. Finally, ten mapped layers for the PML are used.

 \begin{figure}[htbp]
	\centering
	\includegraphics[width=8cm]{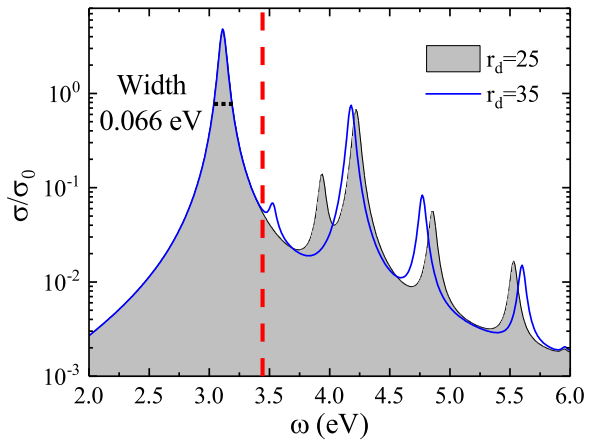}
	\caption{Normalized absorption cross section $\sigma /\sigma_0$ for a jellium nanosphere  with $ N_e=438 $ calculated through the conventional QHT by using two different computation sizes for spill-out electron $r_d$. The black line with grey filling is for $ r_{d}=25a_{0} $ and blue solid line is for a larger size $r_{d}=35a_{0}$. For energy higher than some frequency (indicated by the vertical dashed line), the spectra depend on the computation size $r_d$, which is not stable. The full-width at half maximum for the main LSP resonance is about $0.0658\,eV$, which is nearly equal to the bulk damping rate $\gamma_0=0.066\,eV$. }
	\label{fig2}%
\end{figure}

To show the performance of the conventional QHT, we calculate the normalized absorption cross section ($\sigma/\sigma_0$ ) for a jellium nanosphere with $N_e=438$ electrons. Here, the bulk damping rate $\gamma=\gamma_0=0.066\,eV$ is used for the moment \cite{PhysRevB.93.205405,PhysRevX.11.011049}. The vW coefficient is set to  $\lambda _w=0.4$, which is around the value given in Ref. \cite{PhysRevB.91.115416}. By taking two different computation sizes for the spill-out electron $r_{d}$, we report the results in Fig. \ref{fig2}. There are two main problems. 

One is the computation-size dependent absorption spectra, when the excitation energy is above some critical frequency (indicated by the vertical dashed line in Fig. \ref{fig2}). Different from the TD-DFT spectra (only a shoulder above the main plasmon peak) \cite{PhysRevX.11.011049}, there are some peaks with their position being affected by the computation domain size. These resonances are the analog of Rydberg states for atoms, which are associated with very delocalized states and are numerically affected by the computation domain size \cite{PhysRevB.95.245434,PhysRevB.93.205405}. Physically, for positions far away from the nanosphere, local plasmon frequency $ \omega_p=\sqrt {e^2 n_0/{m_e}{\varepsilon _0}}$ can be much smaller than the excitation $\omega$, due to the exponential decay property of density $n_0$ away from metal surface. The electron-hole pair excitation dominates and large damping (Landau damping due to the interaction between single-particle transitions and surface modes) is expected \cite{RevModPhys.65.677}. By using a density-dependent damping rate $\gamma\propto n_0^{-5/6}$ \cite{Li:15},  the computation becomes stable. However, as pointed out in Ref. \cite{PhysRevB.95.245434}, the induced density results prematurely damped at the metal surface, due to large $\gamma$ near the particle surface, which will naturally have large influence on the LSP resonance. Inspired by the above consideration, we will show that a density-dependent damping exclusively applied in the density-tail region can remove the numerical convergence problem of the QHT, but not affect the main LSP resonance. 

The other problem is the linewidth. By fitting the spectra around the main LSP resonance with a Lorentzian-shaped function, we find that the spectral width is $0.066\,eV$, which is equal to the pre-set value $\gamma_0=0.066\,eV$. Thus, the linewidth broadening is clearly `put in by hand' \cite{Mortensen+2021+2563+2616}, which can not treat the size-dependent broadening for nanostructure of arbitrary shape.  As has been addressed in the GNOR method, introducing the diffusion currents of the conduction electrons can solve this problem. In this work, we will apply this idea to the case for non-uniform ground electron density. 

In the following, we will show how the density-dependent damping can solve the convergence problem and how the size-dependent broadening can be addressed by adding the diffusion electron currents. 
 
\subsection{Convergence problem solved by using a density-dependent damping rate}
Here, we consider the following density-dependent damping rate 
\begin{equation}\label{gaman0}
\gamma \left( r \right) = {\gamma _0}{\left( { {\frac{{{n_+ {e^{ - {r_q}}}}}}{{{n_0}}}} + 1} \right)^{5/6}},
\end{equation}
with $r_q$ a non-negative parameter. $\gamma_0$ is the damping rate in bulk metal. From Eq. \eqref{gaman0}, when ${n_0}\gg n_+ e^{-r_q}$, i.e. inside and much around the metal particle, we have $\gamma \left( r \right)=\gamma_0$. But for ${n_0}\ll n_+ e^{-r_q}$, i.e. in the low electron density region, we have $\gamma \left( r \right)\propto n_0^{-5/6} $, which is similar to that in Ref. \cite{Li:15,PhysRevLett.85.2200}.  

The parameter $r_q$ has a well-defined physical meaning, as it defines where the damping starts to increase rapidly. To simplify the analysis, we adopt the model density $n_0=f_0/(1+e^{k_{mod}(r-R)})$ with $f_0= n_+$ and $k_{mod}=1.0/a_0$, which are much around the values given in Ref. \cite{PhysRevB.93.205405} ($f_0=0.98 n_+$ and $k_{mod}=1.05/a_0$).  Here, $r$ is the distance from sphere center. In this case, Eq. \eqref{gaman0} becomes $\gamma \left( r \right) = {\gamma _0} \left(1+e^{-r_q}+e^{(r-R)/a_0-r_q} \right)^{5/6}$. To further simplify the analysis, let us assume $r_q\geq5$ for the moment, which we will show that $r_q$ should be around $8$ in order to solve the convergence problem. In this case, the damping rate [Eq. \eqref{gaman0}] becomes
\begin{equation*}\label{}
\gamma \left( r \right) = \gamma _0[e^{(r-R)/a_0-r_q} + 1]^{5/6},
\end{equation*}
from which we have $\gamma \left( r \right) \approx 1.78\gamma _0$ at $r=R+r_qa_0$. It is $12.69\gamma _0$ when the position is $(r_q+3)a_0$ away from the metal surface. Then, it grows exponentially with the position further away from the metal surface $\gamma \left( r \right)\approx \gamma _0[e^{(r-R)/a_0-r_q}]^{5/6}$ due to $e^{(r-R)/a_0-r_q}\gg1$, resulting $\gamma \left( r \right)\propto n_0^{-5/6}$. But for positions inside and around the nanosphere, i.e. $r\leq R+(r_q-3)a_0$, it leads to $\gamma \left( r \right) \approx{\gamma _0}$, since $e^{(r-R)/a_0-r_q}\leq e^{-3} \ll 1$. The larger the parameter $r_q$ is, the larger distance from the metal surface the enhanced damping is applied to. Thus, we can conclude that the parameter $r_q$ controls the region where large damping is applied to. 
 
\begin{figure}[htbp]
	\centering
	\includegraphics[width=8cm]{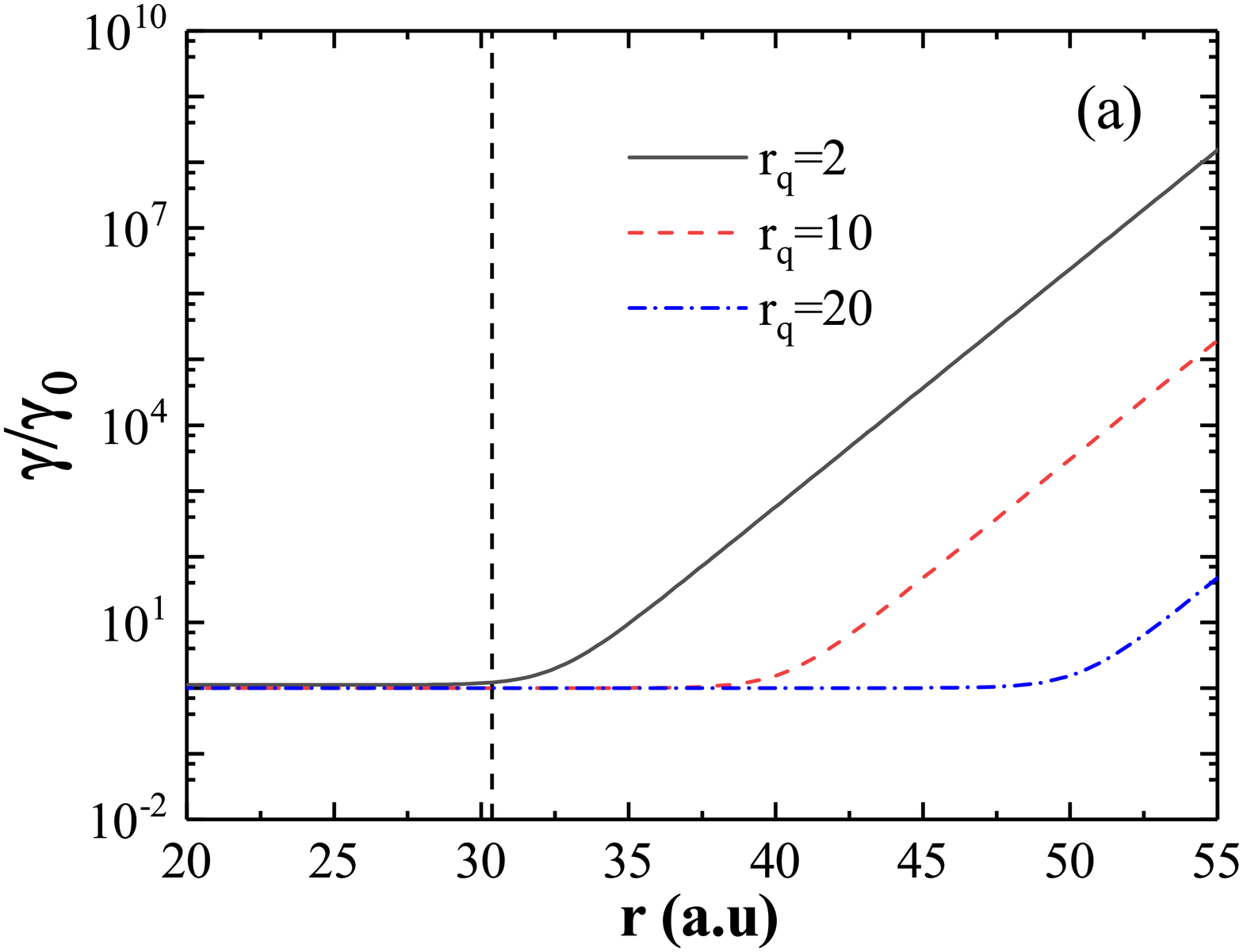}
	\includegraphics[width=8cm]{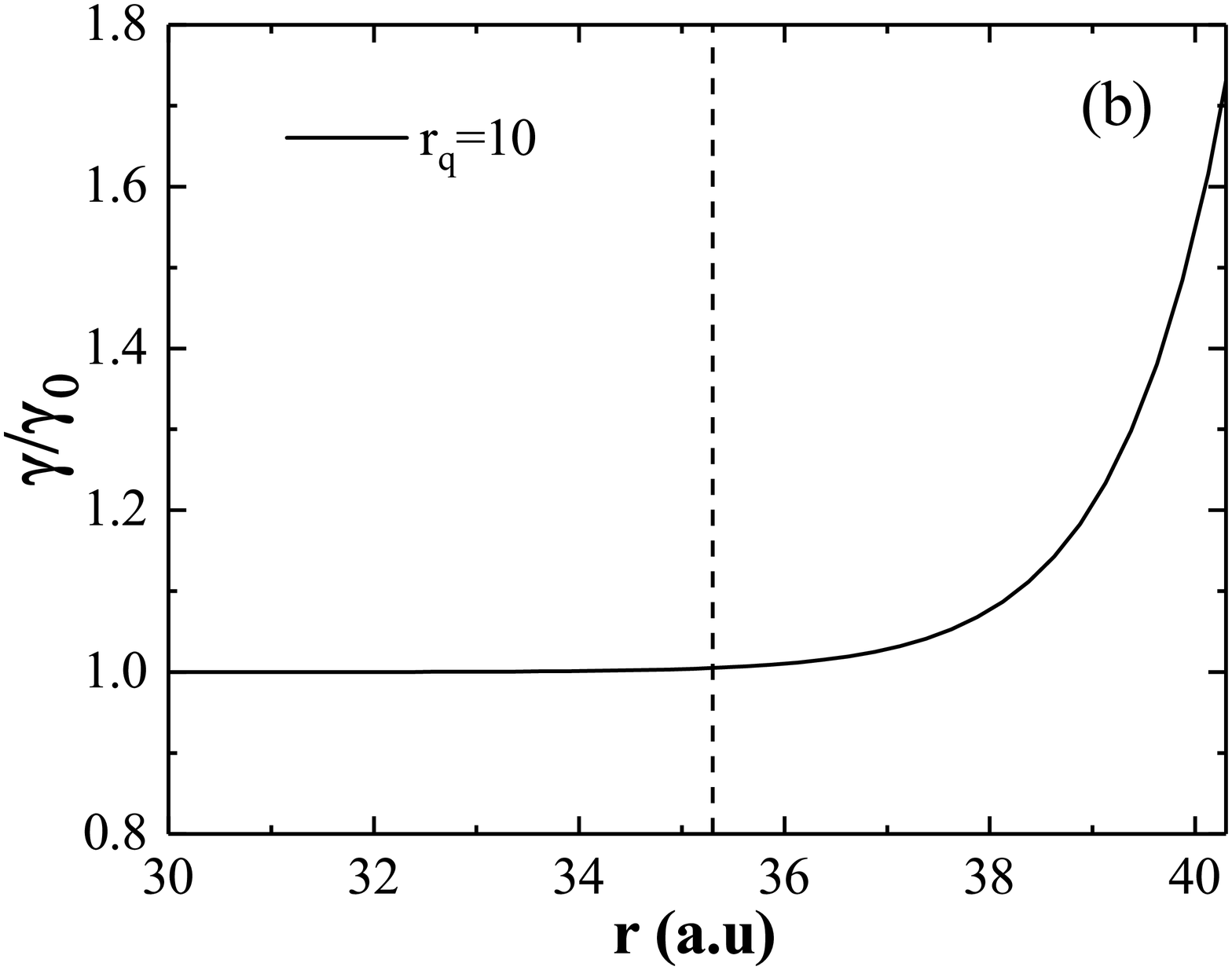}
	\caption{(a) $\gamma/\gamma_0$ by Eq. \eqref{gaman0} with $r_q=2$ (black solid line), $10$ (red dashed line), and $20$ (blue dash-doted line). The vertical dashed line represents the jellium boundary; (b) zoomed view for $r_q=10$ when $R\leq r\leq R+r_qa_0$. Here, the vertical dashed line is at $r=R+(r_q-5)a_0$.}
	\label{fig3}%
\end{figure}

The above properties can be clearly seen from Fig. \ref{fig3}. The vertical line [see Fig. \ref{fig3}(a)] indicates the position of sphere surface. The black solid, red dashed and blue dash-dot lines correspond to $r_q=2$, $10$, and $20$, respectively.  The radius $r$ at which $\gamma/\gamma_0$ starts to increases rapidly is larger for larger $r_q$. Figure 3(b) is a zoomed view for $r_q=10$ with $r$ in the range [$R, R+r_qa_0$], which clearly shows that $\gamma\approx\gamma_0$ for $r\leq R+(r_q-5)a_0$, i.e. on the left of the vertical line in Fig. \ref{fig3}(b).  Since large damping rate $\gamma$ is exclusively applied in the low electron density region ($r\geq R+r_qa_0$), the delocalized states will be efficiently damped. But for the LSP resonance, it is related to the optical response of electrons inside and much close to the metal surface and will not be affected by the large damping in the electron-tail region. This is different from the method by using $\gamma \left( r \right) =\gamma _0(n_+/n_0)^{5/6}$, where the damping starts to increase sharply near the inner surface of the metal and the induced density of the LSP will be prematurely damped. We will numerically show that a small value of $r_q$ can much affect the LSP resonance. 

\begin{figure}[htbp]
	\centering
	\includegraphics[width=8cm]{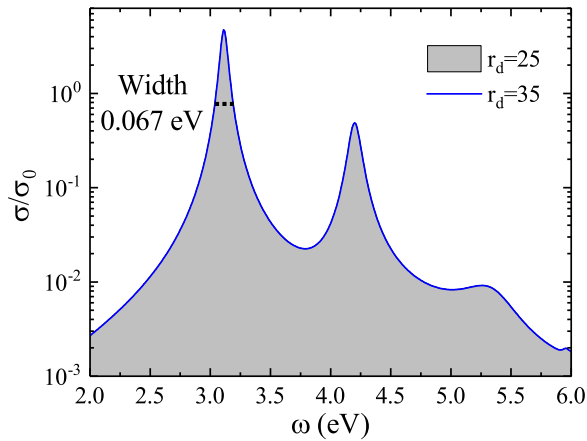}
	\caption{Normalized absorption cross section ($\sigma/\sigma_0$ ) by using the density-dependent damping rate $\gamma$ as defined in Eq. \eqref{gaman0} with other parameters the same as those in Fig. \ref{fig2}. Here, $r_q=10$. The spectra are stable with respect to the computational domain size $r_d$.}
	\label{fig4}%
\end{figure}

By using the above density-dependent damping rate [Eq. \eqref{gaman0}], we find that the solution from QHT becomes stable with respect to the computation size. Figure \ref{fig4} shows the normalized absorption cross section ($\sigma/\sigma_0$ ) by using the density-dependent damping rate $\gamma$ as defined in Eq. \eqref{gaman0} with other parameters the same as those in Fig. \ref{fig2}. Here, we take $r_q=10$ as a demonstration. The absorption spectra by using two different computation sizes (black line with grey filling for $r_d=25a_0$ and blue solid line for $r_d=35a_0$) are the same. In addition, we have checked that this numerical convergence remains as long as the computation domain size for the spill-out electron is about $5a_0$ larger than $r_qa_0$, i.e. $r_d > (r_q+5)a_0$. In this case, it is independent of a special choice for the computation domain size $r_d$. Without using the density-dependent damping rate, see Refs. \cite{PhysRevB.93.205405,PhysRevX.11.011049}, the absorption spectrum is very sensitive to the computation size, where more and more modes appear (and with reduced intensities) with increasing $r_d$. Thus, the convergence problem can be solved by using the density-dependent damping rate [Eq. \eqref{gaman0}]. 

\begin{figure}[htbp]
	\centering
	\includegraphics[width=8cm]{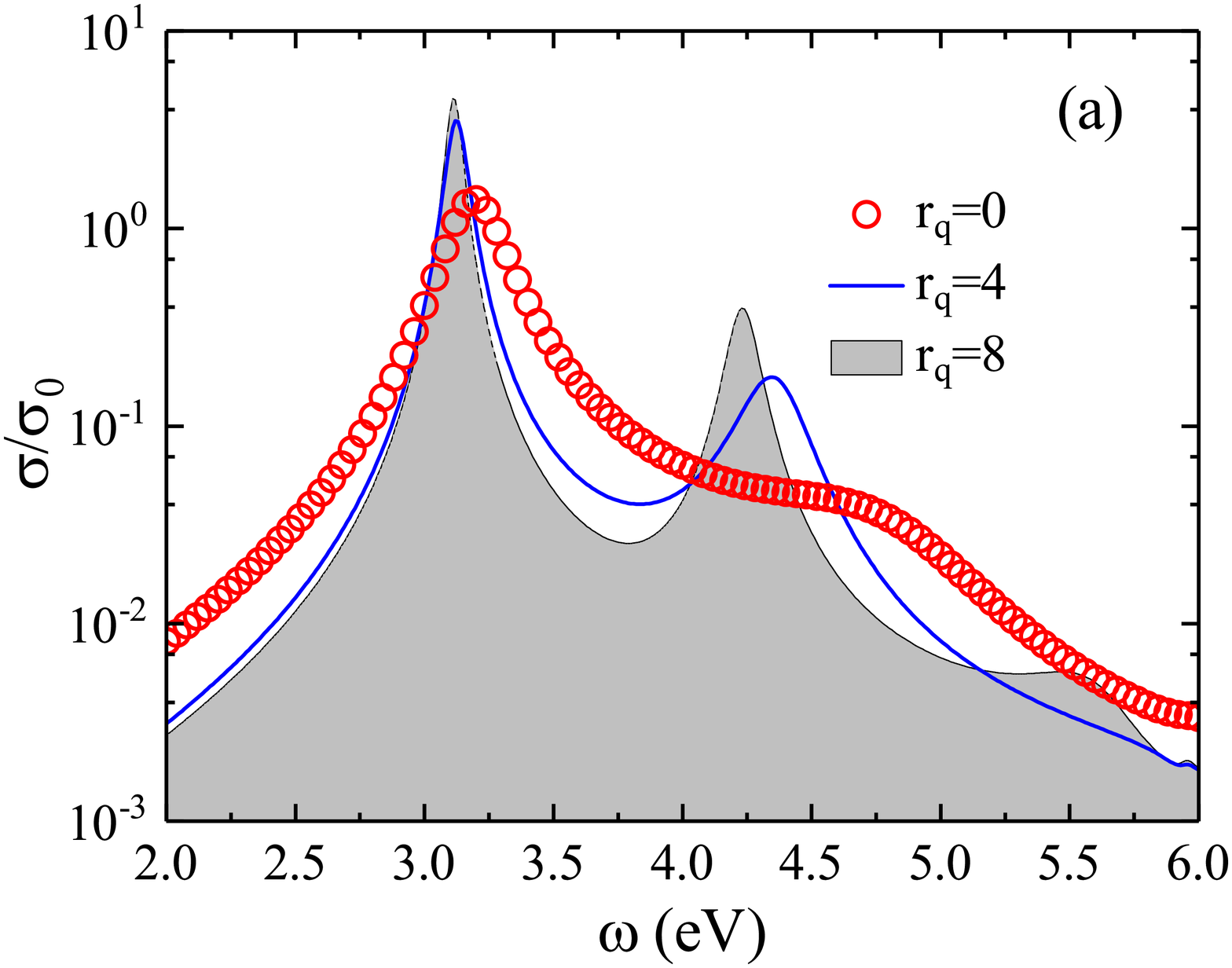}
	\includegraphics[width=8cm]{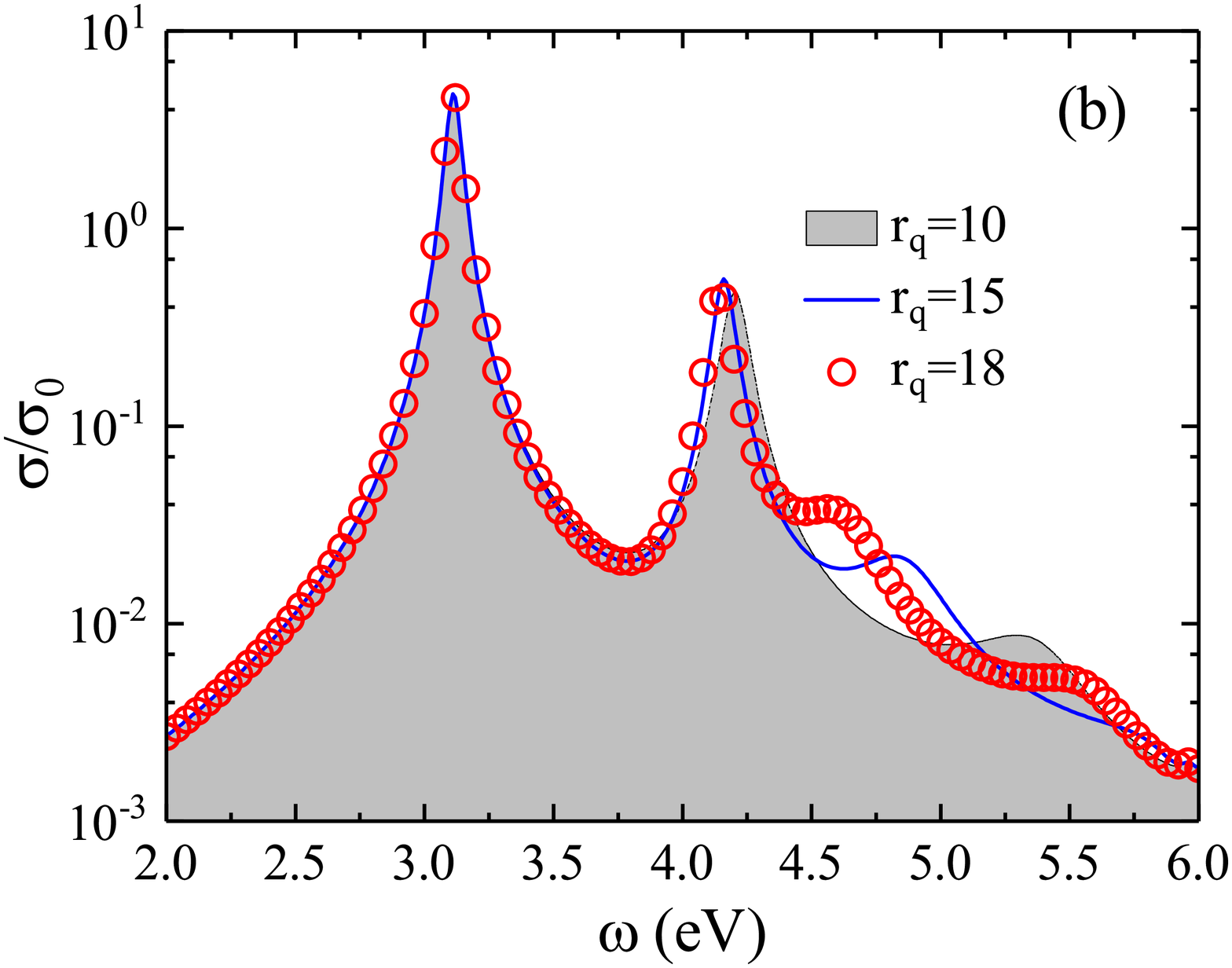}
	\caption{Effect of different parameters $r_q$ on the normalized absorption cross section for a jellium sodium nanosphere with $ N_{e}=438 $. (a) small $r_q$, with $r_q=0$ (red circle), $4$ (blue solid line), and $8$ (black solid line with grey filling); Too small $r_q$ affects much the main LSP resonance. (b) large $r_q$, with $r_q=10$ (black line with grey filling), $15$ (blue solid line) and $18$ (red circle). Sufficiently large $r_q$ leads to stable main LSP peak and the second peak.}
	\label{fig5}
\end{figure}

Another important aspect is how to choose the parameter $r_q$, since it controls the position where large damping is applied. Figure \ref{fig5} shows the normalized absorption cross section $ {\sigma }/{{{\sigma }_{0}}}$ for different $r_q$. See Fig. \ref{fig5}(a), when $r_q$ is very small, i.e.  $r_q=0$, the main LSP resonance energy is higher and the linewidth is larger than those for $r_q=8$. They are consistent with the previous description where large damping ($\gamma \propto n_0^{-5/6}$) applied from the inner of nanosphere leads to a blue shift and a large broadening of the LSP resonance \cite{Li:15}. However, for even larger $r_q$ [see Fig. \ref{fig5}(b) for $r_q=10$, $15$, and $18$], the main LSP spectra become stable. In addition, the linewidth is nearly the same as that in Fig. \ref{fig2} obtained by using $\gamma=\gamma_0$. Thus, a stable main LSP spectra can be obtained by using the density-dependent damping rate [Eq. \eqref{gaman0}] with sufficiently large $r_q$, i.e. $r_q\geq8$. Similarly, only in this case can the second peak be stable.  See Fig. \ref{fig5}(a), for the peak around $4.2\,eV$, large difference can be seen from the curves for $r_q=0$, $4$, and $8$. Differently, for sufficiently large $r_q$ [see Fig. \ref{fig5}(b)], the second peak becomes stable and it is also nearly independent of the parameter $r_q$. We have checked that this property remains when $r_q=40$ and $80$. By taking sufficiently large $r_q$, both the main LSP peak and the second peak are stable. 

However, the larger the $r_q$ is, the more peaks appear in the spectra at high energy (above the main LSP). See Fig. \ref{fig5}(a), there is only one high-energy peak for $r_q=0$ and $4$, but two for $r_q=8$. For the curves in Fig. \ref{fig5}(b), three high-energy peaks can be clearly seen for $r_q=15$ and $18$. When $r_q$ is extremely large, i.e. $r_q\to+\infty$,  Eq. \eqref{gaman0} becomes $\gamma\to\gamma_0$, leading to the conventional QHT where $\gamma=\gamma_0$ is used. In this case, an infinite number of peaks should appear with an infinite computational domain size \cite{PhysRevX.11.011049}. By calculating the Feibelman $d$-parameter, we find that these peaks are related to the Bennett states, since the real part experiences an abrupt change from the positive to negative value, while its imaginary part shows a peak. These Bennett states are the analog of Rydberg states for atoms \cite{PhysRevB.95.245434}, where higher order modes are more extended in space. Thus, for larger $r_q$, more peaks remain except for the modes with much higher order, since large damping is applied only in the region far away from the metal surface ($r>R+r_qa_0$). As stated in the introduction, there should be large Landau damping in the low electron density region and $r_q$ can not be too large.

From the above results, we see that the parameter $r_q$ should be sufficiently large,  i.e. $r_q\geq8$, but can not be too large. As pointed out in Ref. \cite{BROWN1974489}, Coulomb repulsion effects might lead to a tendency to Wigner lattice formation in the electron density tail region. For uniform electron gas, an estimate that $r_s>40a_0$ according to Lindemann criterion or $r_s=106a_0$ \cite{ceperley1980ground} by quantum Monte Carlo simulation is required to give a stable Wigner crystal. Here, we find that the position for $r_s=40a_0$ and $r_s=106a_0$ are located at $r=R+5.5a_0$ and $r=R+7.3a_0$, respectively. In this case, a reasonable value for $r_q$ should not be much larger than $6$ (around $5.5$ and $7.3$). In the following without otherwise statement, we will use the density-dependent damping rate [Eq. \eqref{gaman0}] in place of the bulk term $\gamma_0$. The parameter $r_q=10$ is used to ensure a stable solution for the main LSP resonance peak and the first Bennett state. 

It should be noted that the damping rate described by Eq. \eqref{gaman0} can be applied to nanoparticle of arbitrary shape. Since the ground density shows similar exponential decay in the electron tail region, i.e. $n_0=b_Qe^{-k_Q x} $ with $x$ being the distance away from the metal surface \cite{PhysRevB.91.115416}, it is the same as the above model density in the electron tail region and similar analysis can be made. In this work, the above density-dependent damping rate [Eq. \eqref{gaman0}] will be applied to the case of a nanorod and it is found that it works well.  

\subsection{Width of the absorption spectra resolved by density-dependent diffusion}

In the previous section, we have shown that the convergence problem can be solved by using a density-dependent damping rate [Eq. \eqref{gaman0} with $r_q$ around $10$]. However, the width of the main LSP spectra is nearly equal to the input damping rate $\gamma_0$ and it is hard to add a Kreibig term for nanostructrue of nonspherical shape. In this subsection, we attempt to solve the size-dependent broadening by introducing the diffusion electron current. Following Ref. \cite{Mortensen2014A}, a current density $eD\nabla n_1=D\nabla (\nabla\cdot \mathbf{P})$ due to diffusion should be added and the constitutive equation [Eq. \eqref{subeq:2}] becomes
\begin{eqnarray}\label{PEequation}
&&\frac{{e{n_0}}}{{{m_e}}}\nabla {\left( {\frac{{\delta G\left[ n \right]}}{{\delta n}}} \right)_1} + \left( {{\omega ^2} + i\gamma \left[ {{n_0}} \right]\omega } \right){\bf{P}}  \nonumber\\
&&+{\left( {\gamma \left[ {{n_0}} \right] - i\omega } \right)D} \nabla \left( {\nabla  \cdot \mathbf{P}} \right) =  - {\varepsilon _0}\omega _p^2({{\bf{E}}_i} + {{\bf{E}}_s}).
\end{eqnarray}
Here, $D$ is the diffusion coefficient. Under the GNOR model with uniform ground electron density $n_0$, the relation $D=3\sqrt{10}A_1v_F^2/5\omega_p\propto n_0^{1/6}$ provides an accurate prediction for the linewidth broadening \cite{Raza_2015}. In an attempt to generalize this result to the case for nonuniform ground density $n_0$, one can assume a similar expression 
\begin{equation}\label{densityDifu}
D = A\frac{v_F^2}{{{\omega _p}}},
\end{equation}
with the coefficient $A$ to be determined. In this case, the diffusion coefficient shows a weak density-dependent behavior, i.e. $D\propto n_0^{1/6}$. 

\subsection{Numerical determination of the coefficients $\lambda_w$ and $A$}

\begin{figure}[htbp]
	\centering
	\includegraphics[width=8cm]{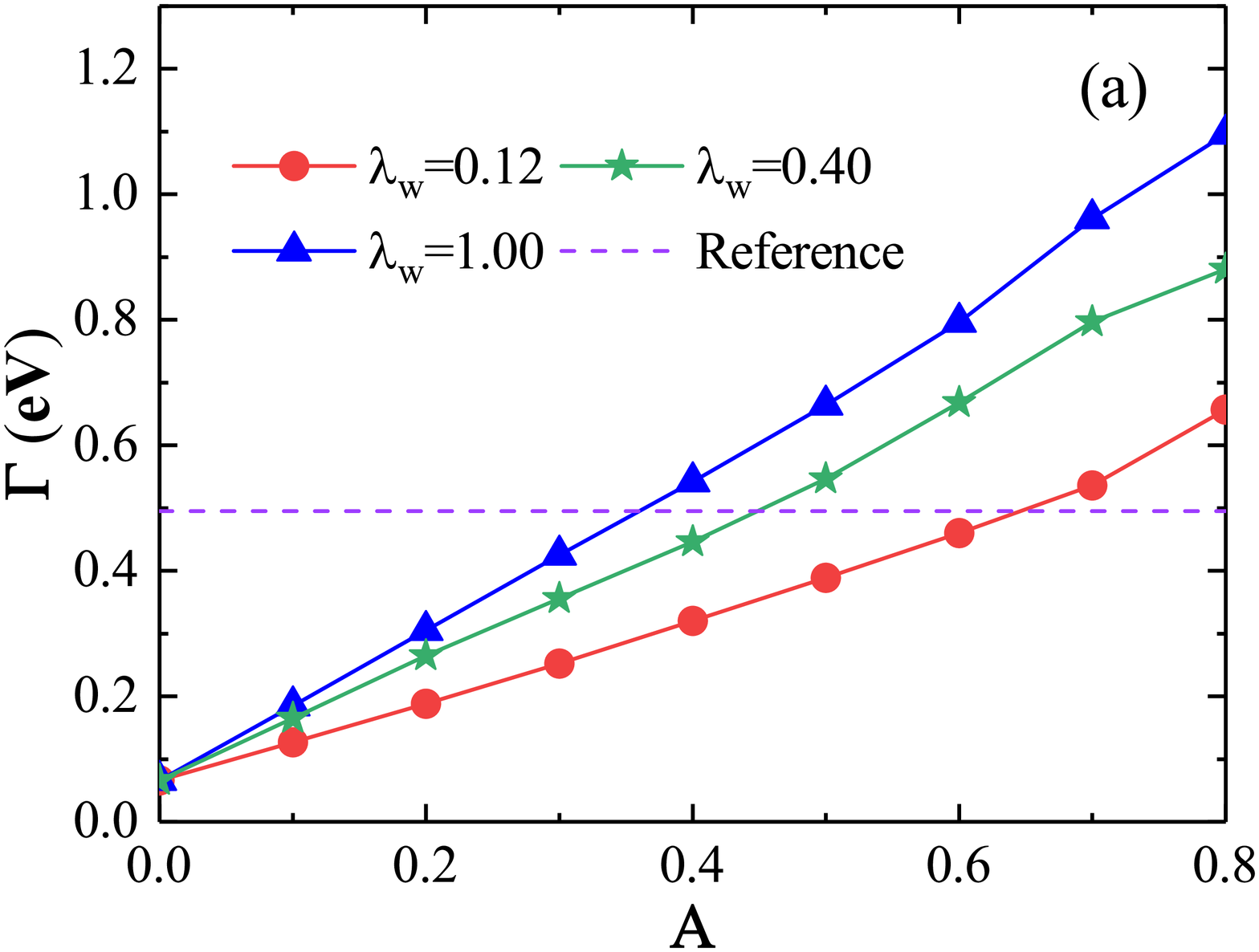}
	\includegraphics[width=8cm]{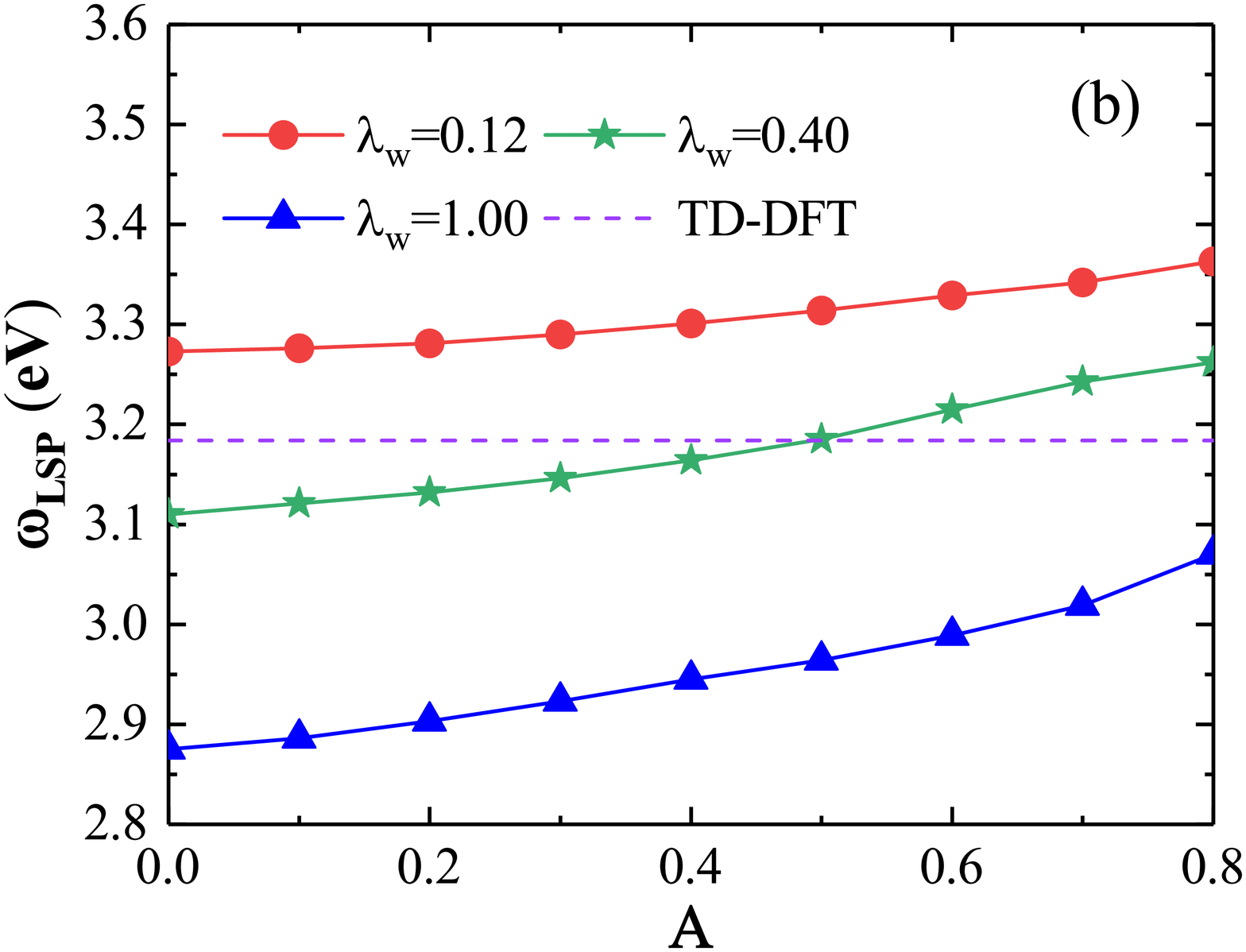}
	\caption{(a) Width $\Gamma$, and (b) resonance energy $\omega_{LSP}$ for the main LSP as a function of the coefficient $A$. Here, we consider three typical values of $\lambda_{w}$, i.e. $\lambda_{w}=0.12$, $0.40$, and $1.00$. The purple horizontal dashed lines in (a) and (b) represent the reference results from Kreibig formula $\Gamma=\gamma_0+v_F/R$ and by TD-DFT \cite{PhysRevB.93.205405}, respectively. Here, $ N_e=438 $ ($R=1.61\,nm$). With increasing $A$, $\Gamma$ increases quickly, while $\omega_{LSP}$ increases much slowly.}
	\label{fig6}%
\end{figure}

In the present form of the QHT [coupled Eqs. \eqref{subeq:1} and \eqref{PEequation}], the two important coefficients ${{\lambda }_{w}}$ and $A$ should be determined in order to give not only the correct  main LSP resonance energy $\omega_{LSP}$ but also the spectra width $\Gamma$. Figure \ref{fig6} shows how the coefficient $A$ affects the width $\Gamma$ and the resonance energy $\omega_{LSP}$ when three typical values of $\lambda_{w}$ are used, i.e. $\lambda_{w}=0.12$, $0.40$, and $1.00$. For the width [see Fig. \ref{fig6}(a)], when $A=0$, we have $\Gamma\approx\gamma_0$ for all three values of $\lambda_{w}$. With increasing $A$, the width $\Gamma$ increases linearly with a very large slope. For example, the slopes are about $0.712\,eV$, $1.025\,eV$, and $1.278\,eV$ when $\lambda_{w}=0.12$, $0.40$, and $1.00$, respectively. The width $\Gamma$ increases quickly with increasing $A$. The horizontal dashed line represents the value by Kreibig formula  $\Gamma=\gamma_0+v_F/R$, which can be considered as a reference. It intersects with the curves for $\lambda_{w}=0.12$, $0.40$, and $1.00$ at $A=0.65$, $0.45$, and $0.36$, respectively.
Thus, in order to obtain the required width, the coefficient $A$ should be in the range $[0.36,0.65]$, with a mid-value around $0.5$ for the nanosphere investigated ($N_e=438$). 

But for the resonance energy $\omega_{LSP}$ [see Fig. \ref{fig6}(b)], it increases much slowly with increasing $A$. For example, when $A$ changes from $0$ to $0.8$, the total variations for $\omega_{LSP}$ are about $0.090\,eV$, $0.152\,eV$, and $0.195\,eV$ for $\lambda_{w}=0.12$, $0.40$, and $1.00$, respectively.  The three curves in Fig. \ref{fig6}(b) are nearly equidistant with a large separation. A smaller $\lambda_{w}$ leads to a much higher resonance energy $\omega_{LSP}$.  The horizontal dashed line represents the result by TD-DFT, which only intersects with the curve for $\lambda_{w}=0.4$ at $A=0.5$. Too large or too small value for $\lambda_{w}$ can not give the required resonance energy, i.e. the resonance energies $\omega_{LSP}$ by using $\lambda_{w}=0.12$ and $1.00$ are either much higher or lower than the reference value. These results show that $\lambda_{w}$ should be around $0.4$. In this case, see Fig. \ref{fig6}(a), the coefficient $A$ should be much around $0.45$ (the intersection point between the curve for $\lambda_{w}=0.40$ and the reference horizontal dashed line), which further confirm that $A$ should be around $0.5$.

As is shown above, with increasing $A$, the width $\Gamma$ increases quickly, while $\omega_{LSP}$ increases very slowly. In order to give the required $\omega_{LSP}$ and $\Gamma$, the coefficients $A$ and $\lambda_{w}$ should be around $0.5$ and $0.4$, respectively. Compared with $A$, the coefficient $\lambda_w$ has less influence on the width $\Gamma$, but more on the resonance energy $\omega_{LSP}$.

\begin{figure}[htbp]
	\centering
	\includegraphics[width=8cm]{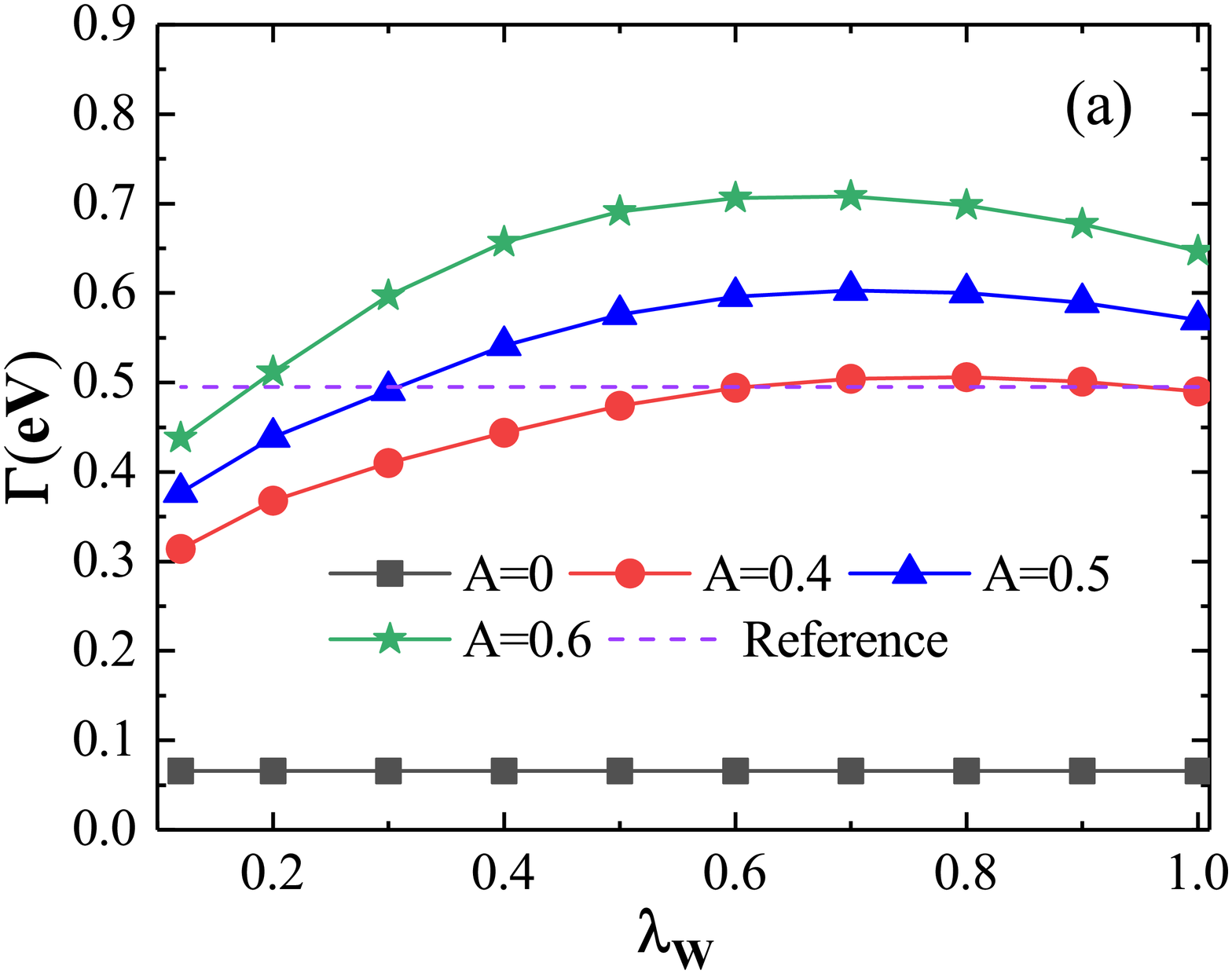}
	\includegraphics[width=8cm]{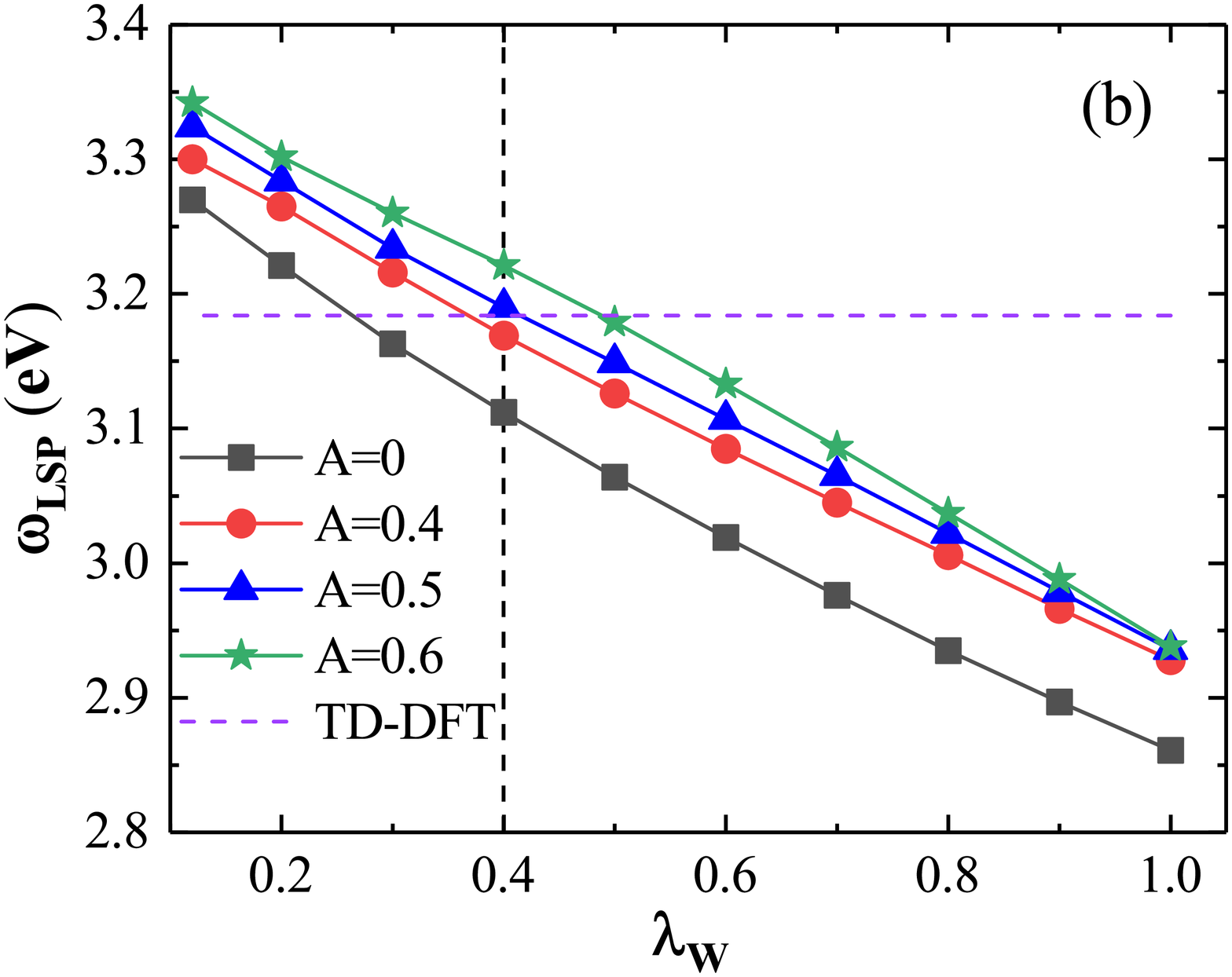}
	\caption{(a) Width $\Gamma$, and (b) resonance energy $\omega_{LSP}$ for the main LSP as a function of the vW coefficient $\lambda_w$. Here, besides for $A=0$, we consider three values of $A$ around the required value $0.5$, i.e. $A=0.4$, $0.5$, and $0.6$. The purple horizontal dashed lines in (a) and (b) represent the reference results from Kreibig formula $\Gamma=\gamma_0+v_F/R$ and by TD-DFT \cite{PhysRevB.93.205405}, respectively. $N_e=438 $ ($R=1.61\,nm$). $\Gamma$ varies slowly, while $\omega_{LSP}$ decreases quickly with increasing $\lambda_w$. The vertical line in (b) is located at $\lambda_w= 0.4$.}
	\label{fig7}%
\end{figure}

To see this more clearly, we plot the spectral width as a function of $\lambda_w$ in Fig. \ref{fig7}(a). The coefficient $A$ is around $0.5$, i.e. $A=0.4$, $0.5$ and $0.6$, in sharp contrast with $A=0$. The horizontal dashed line located at $0.495\,eV$ represents the reference result by the Kreibig formula. The three curves for $A=0.4$, $0.5$ and $0.6$ are nearly equidistant and they vary very slowly over a wide range of $\lambda_{w}$ ($0.12\leq\lambda_w\leq 1.00$). For example, when $A=0.4$, the width ranges from $0.314\,eV$ to about $0.506\,eV$, with its maximum nearly being equal to the reference value $0.495\,eV$. While for $A=0.6$, the width falls between $0.438\,eV$ and $0.708\,eV$, with a minimum close to the reference. When $A$ is much less than $0.4$ or much larger than $0.6$, the required width can not be obtained. It is further shown that the coefficient $A$ should be around $0.5$ in order to give the required spectral broadening, i.e. in the range $ 0.4<A<0.6$. Compared with the results in Fig. \ref{fig6}(a) where the width $\Gamma$ increases quickly with increasing $A$, the coefficients $\lambda_w$ has less influence on the width $\Gamma$.

Figure. \ref{fig7}(b) shows the main LSP resonance energy $\omega_{LSP}$ as a function of the coefficient $\lambda_w$. The four curves for $A=0$, $0.4$, $0.5$, and $0.6$ are nearly equidistant straight lines. The slopes are very large with an average value around $-0.44\,eV$, which means a quick decrease for $\omega_{LSP}$ with increasing $\lambda_w$. For example, see $A=0.4$ in Fig. \ref{fig7}(b), $\omega_{LSP}$ decreases linearly from $3.300\,eV$ to $2.928\,eV$, with a large variation about $0.372\,eV$, when the coefficient $\lambda_w$ increases from $0.12$ to $1.00$. This large red-shift with increasing $\lambda_w$ is mainly due to the electron density spill-out in free space \cite{toscano2015resonance,PhysRevB.74.165421}, with the low value of $\lambda_w$ corresponding to less spill-out \cite{Li:15}. 

Different from the above remarkable effect of $\lambda_w$ on the resonance energy, the coefficient $A$ has small effect. Compared with the case for $A=0$, i.e. without taking into account the diffusion effect, a larger $A$ leads to higher resonance energy. However, the blue shift for $\omega_{LSP}$ is rather small with increasing $A$. For example, at $\lambda_w=0.40$ [indicated by the vertical line in Fig. \ref{fig7}(b)], the main LSP resonance energies are $3.112\,eV$, $3.169\,eV$, $3.190\,eV$ and $3.221\,eV$, when $A=0$, $0.4$, $0.5$, and $0.6$, respectively. Only a small difference is found, i.e. a maximum difference about $0.052\,eV$, when the coefficient $A$ varies in the required range ($0.4<A<0.6 $).  

In Fig. \ref{fig7}(b), the horizontal dashed line located at $3.184\,eV$ represents the TD-DFT result \cite{PhysRevB.93.205405}, which can be considered as a reference. It intersects with the curves for $A=0.4$ and $A=0.6$ at $\lambda_w=0.35$ and $\lambda_w=0.47$, respectively. Thus, the required coefficient $\lambda_w$ should be in the range $ 0.35\leq \lambda_w \leq 0.47$. It is interesting to see that the value $\lambda_w=0.435$, required to give the work function close to the DFT value for semi-infinite metal (sodium) \cite{PhysRevB.91.115416}, is in this range. Only for $ 0.35\leq \lambda_w \leq 0.47$, can the main LSP resonance energy by QHT agree well with that by TD-DFT for the present nanosphere. Thus, we fix $\lambda_w=0.4$, which is in the allowed range and around the median value. In this case, from Fig. \ref{fig7}(a) at $\lambda_w=0.4$, one can see that the purple horizontal dashed line (reference value) is almost in the middle of red circle ($A=0.4$) and blue triangle ($A=0.5$), which further proves $0.4<A<0.5$.

\begin{figure}[htbp]
	\centering
	\includegraphics[width=8cm]{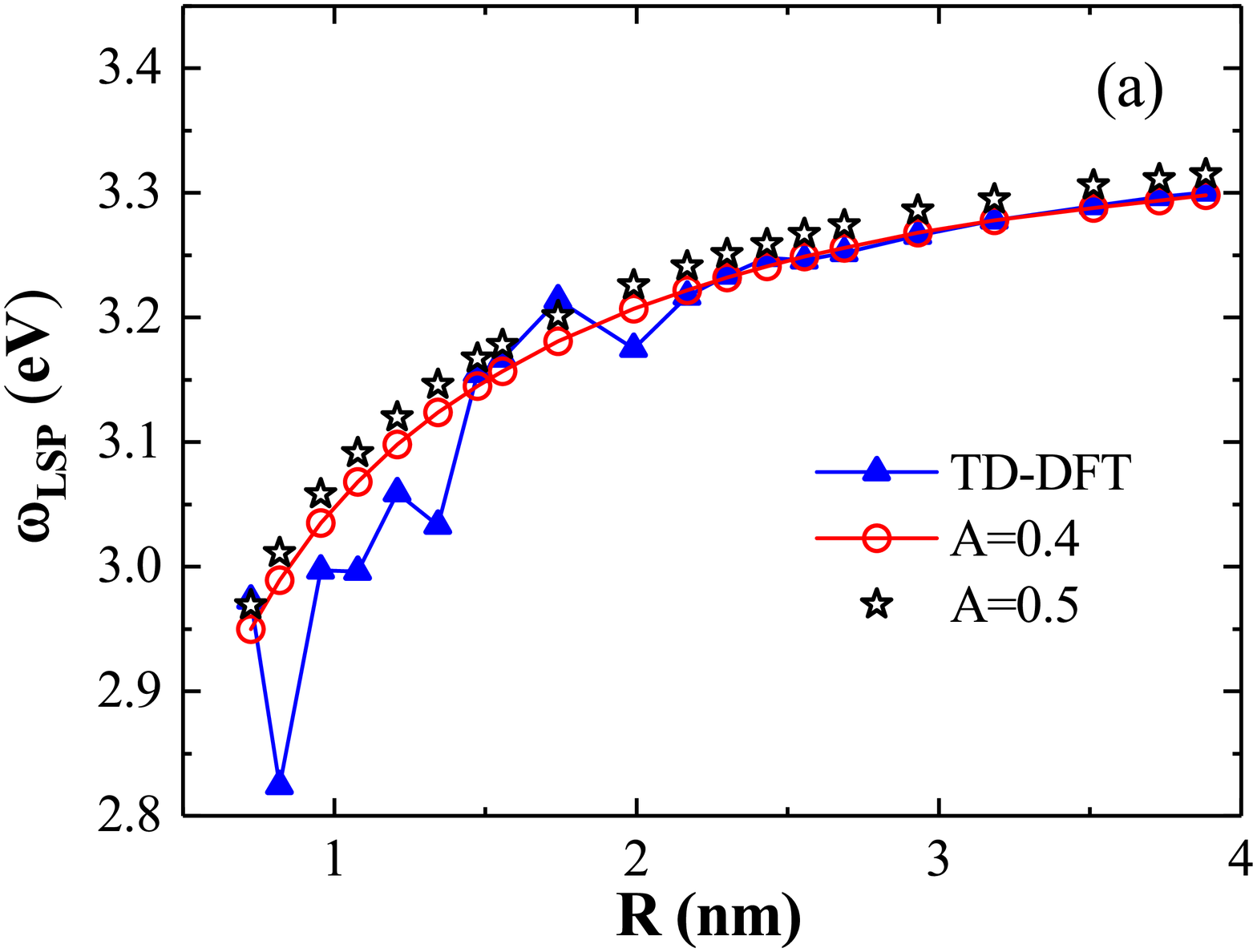}
	\includegraphics[width=8cm]{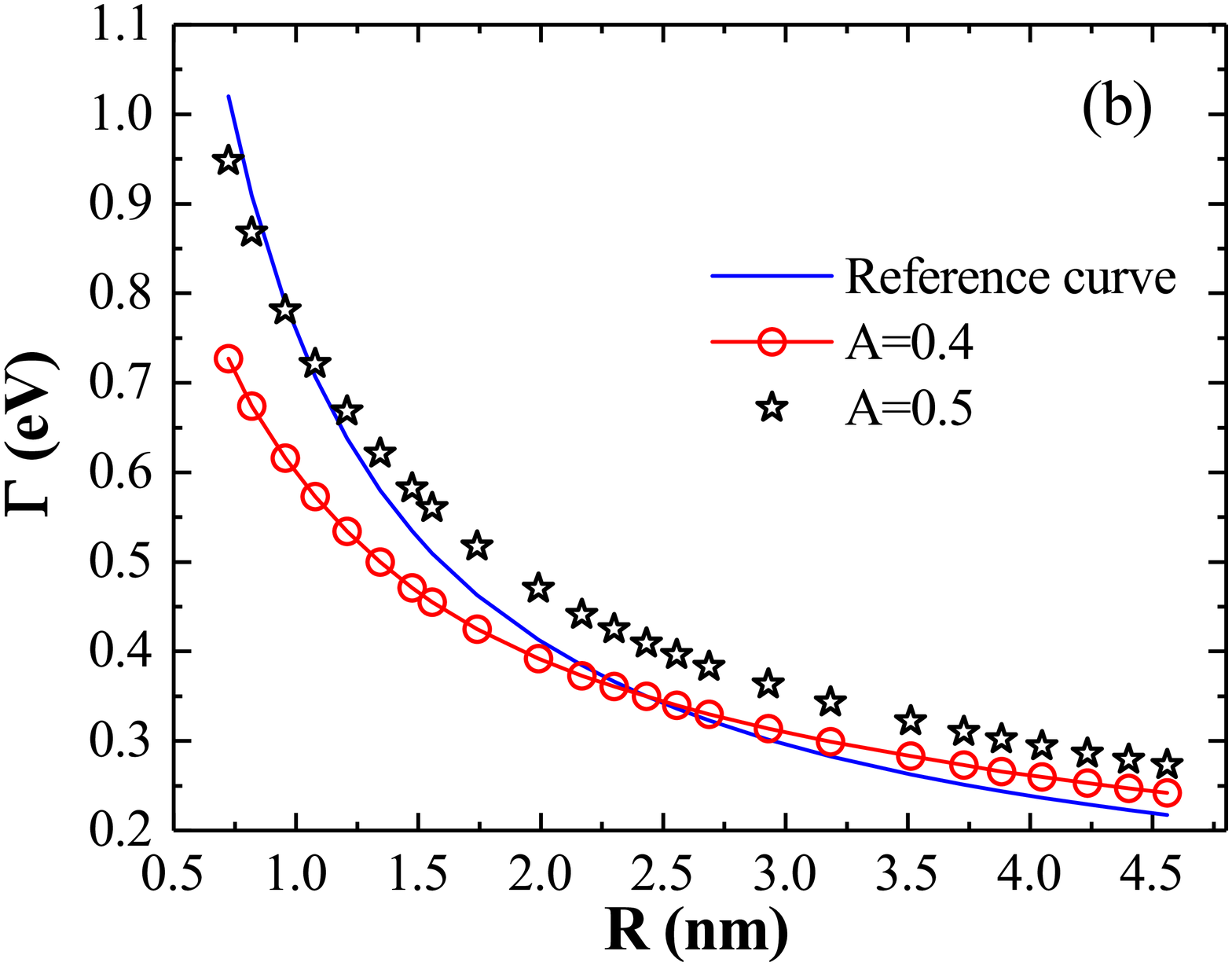}
	\caption{(a) Resonance energy $\omega_{LSP}$, and (b) width $\Gamma$ for the main LSP for nanosphere of various sizes. Here,  $A=0.4$ (red circles) and $A=0.5$ (black stars) are used. The coefficient $\lambda_w$ is fixed to $0.4$. In (a), the blue triangles are the reference results by TD-DFT \cite{PhysRevB.93.205405,PhysRevX.11.011049}. The black stars ($A=0.5$) are a little above the red circles ($A=0.4$). The results by $A=0.4$ show excellent agreement with the TD-DFT value. In (b), the reference curve is from Kreibig formula $\Gamma=\gamma_0+v_F/R$.}
	\label{fig8}%
\end{figure}

As stated above, both the main LSP resonance energy $\omega_{LSP}$ and the width $\Gamma$ can agree well with the reference results by using $\lambda_w=0.4$ and $0.4 < A < 0.5$ for a nanosphere with $ N_e=438 $ ($R=1.61\,nm$). Applying to nanosphere with various radius ranging from $0.72\,nm$ ($Ne=40$) to $3.88\,nm$ ($Ne=6174$), we plot the results in Fig. \ref{fig8}. Here, we take two different values for coefficient $A$, i.e. $A=0.4$ and $0.5$. $\lambda_w$ is fixed to $0.4$. For the resonance energy $\omega_{LSP}$, see Fig. \ref{fig8}(a), compared with $A=0.4$, the results for $A=0.5$ are a little larger (within $0.025\,eV$). Moreover, their differences are nearly independent of the sphere radius. Note that the results for $A=0.4$ (red circle) are more accurate than those for $A=0.5$. They show almost exactly the mean trajectory of TD-DFT data. When $R\geq 2.16\,nm$ ($N_e\geq1074$ ), excellent agreement with the TD-DFT can be obtained. From these results, one can conclude that $\omega_{LSP}$ can be obtained by $A=0.4$ or $A=0.5$ for nanosphere of various sizes, especially for larger nanosphere. In addition, $A=0.4$ is more proper than $A=0.5$.

To see more clearly, we show the mean average errors (MAE) for the main LSP resonance energy with respect to reference TD-DFT in Table \ref{table1}. Here, we consider two sets of nanospheres. The first one contains thirteen nanospheres with $338 \leq N_e \leq 5470$, and the second contains nine nanospheres by excluding the four smallest nanosphers in the first one, where $1074 \leq N_e \leq 5470$. For the first set, see the middle column with $ N_e\geq 338 $, MAE is about $19.5\,meV$ with $A=0.5$, which is a little larger than $9.1\,meV$ obtained by $A=0.4$. For the second set, see the last column with $ N_e\geq 1074$,  MAE by $A=0.5$ shows little decrease compared with that in the middle column. However, in this case, MAE by $A=0.4$ decrease to a very low value, i.e. about $3.7\,meV$, which means that $A=0.4$ provides highly accurate resonance energy. It should be noted that MAE for $\omega_{LSP}$ by both $A=0.4$ and $A=0.5$ are smaller than those by previous methods (see OF9QHT9 and KSQHT1 in Table \ref{table1}), especially for $A=0.4$. These results further prove that $\lambda_w=0.4$ is appropriate, since the resonance energy is mainly determined by this coefficient. 

Figure \ref{fig8}(b) shows the width $\Gamma$ for various nanospheres. Neither $A=0.4$ (red circles) nor $A=0.5$ (black stars) can provide an overall agreement with the reference values. MAE for the width $\Gamma$ over $0.72\,nm\leq R \leq 4.56\,nm$ ($40\leq N_e \leq 10^4$) are $58.8\,meV$ and $51.4\,meV$ when $A=0.4$ and $A=0.5$, respectively. For smaller nanospheres, i.e. $R\leq 1.56\,nm$ ($N_e\leq 398$), the results by $A=0.4$ deviate more from the reference values than those by $A=0.5$. However, on the other hand for larger nanospheres, the results by $A=0.4$ is much closer to the reference value. The above results mean that larger coefficient $A$, i.e. $A>0.4$, is required for smaller nanosphere. On the other hand, a smaller value , i.e. $A<0.4$, is needed for larger nanosphere. Thus, in order to obtain more accurate width $\Gamma$, $A$ should dependent on sphere radius, with smaller $R$ requiring a little larger $A$.

\begin{table}[h]
	\caption{\label{table1}%
		Performance of the QHT approach using different values of coefficients $A$, $\lambda_w$, and input density. The mean average errors (MAE) for the main LSP resonance energy with respect to reference TD-DFT are shown. Here, we consider two sets of nanospheres. The first one contains thirteen nanospheres with $338 \leq N_e \leq 5470$ (the middle column), and the second contains nine nanospheres by excluding the four smallest nanosphers in the first one, where $1074 \leq N_e \leq 5470$ (the last column). The first three rows are results by QHT with $\lambda_w=0.4$. Results for the last three rows and TD-DFT are from supplemental material of Ref. \cite{PhysRevB.93.205405}. The OF9QHT9 represents self-consistent QHT with $\lambda_w=1/9$ for both ground and excited state. The KSQHT1 and ModQHT1 are the methods with the ground density obtained by KS DFT calculation and an analytical model input density for sphere, while $\lambda_w=1$ is used for the excited state.  Here, the unit is $meV$.}
	
	\begin{ruledtabular}
		\begin{tabular}{lcc}
			\textrm{Method}&
			\textrm{MAE($N_e \geq 338$)}&
		    {\textrm{MAE($N_e \geq 1074$)}}\\
			\colrule
			A=0.5 & 
			19.5 & 18.6 \\
		    A=0.4 & 
		    9.1 & 3.7 \\
			A from Eq. \eqref{fitA} & 
			\textbf{7.2} & \textbf{2.9} \\
			OF9QHT9  & 
			40.5 & 32.7 \\
			KSQHT1  & 
			23.2 & 21.3 \\
			ModQHT1  & 
			11.5 & 5.0 \\
		\end{tabular}
	\end{ruledtabular}
\end{table}

\begin{figure}[htbp]
	\centering
	\includegraphics[width=8cm]{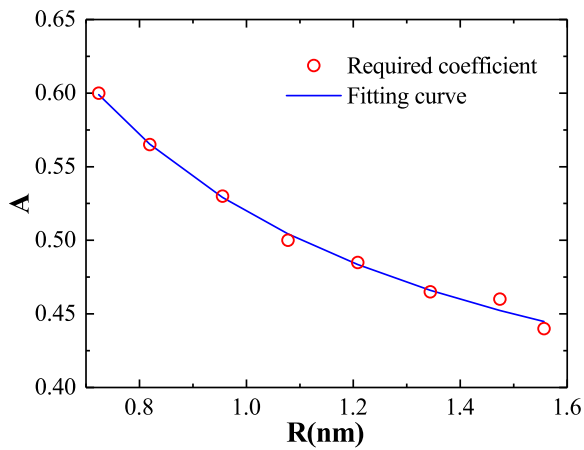}
	\caption{The required coefficient $A$ as a function of the nanosphere radius.  An excellent fitting function [see Eq. \eqref{fitAR}] is obtained. Note that nanosphere radii are small, ranging from $0.72\,nm$ to $1.56\,nm$ ($40 \leq N_e \leq 398$). }
	\label{fig9}%
\end{figure}

As stated above, we expect a size dependent coefficient $A$ in order to give the required width. To this end, we then follow a similar procedure presented in Ref. \cite{Raza:15}. The coefficient $A$ is varied until the width of spectra agrees with the Kreibig formula. This procedure is repeated for eight nanospheres with radius ranging from $0.72\,nm$ to $1.56\,nm$ ($N_e=40,\,58,\,92,\,132,\,186,\,256,\,338,\,398$). In Fig. \ref{fig9}, we plot the required coefficient $A$ as a function of the sphere radius $R$ (see red circles). It decays from $0.60$ to $0.44$ when $R$ increases from $0.72\,nm$ to $1.56\,nm$, which further proves that smaller nanosphere needs larger coefficient $A$. By fitting these results, we find that
\begin{equation}\label{fitAR}
A=0.31+\frac{3.94a_0}{R}.
\end{equation}
Since $R=N_e^{1/3}r_s$ with $r_s=4a_0$ for jellium nanosphere, the above equation can also be written as
\begin{equation}\label{fitA}
A=0.31+\frac{0.99}{N_e^{1/3}}, 
\end{equation}
which shows a weak size dependent behaviour. Although Eqs. \eqref{fitAR} and \eqref{fitA} are equivalent for nanosphere, Eq. \eqref{fitA} can be directly applied to nonspherical nanostructure since it depends on the total electron number $N_e$, while Eq. \eqref{fitAR} can not be. For  nanostructure of nonspherical shape, the radius $R$  should be replaced by some `effective length' $R_{eff}$ \cite{doi:10.1063/1.445794} if one persists to use Eq. \eqref{fitAR}. In section \ref{section3}, we will give several possible $R_{eff}$ for nanorod and show that there is minor difference for the predicted resonance energy and width. 

\begin{figure}[htbp]
	\centering
	\includegraphics[width=8cm]{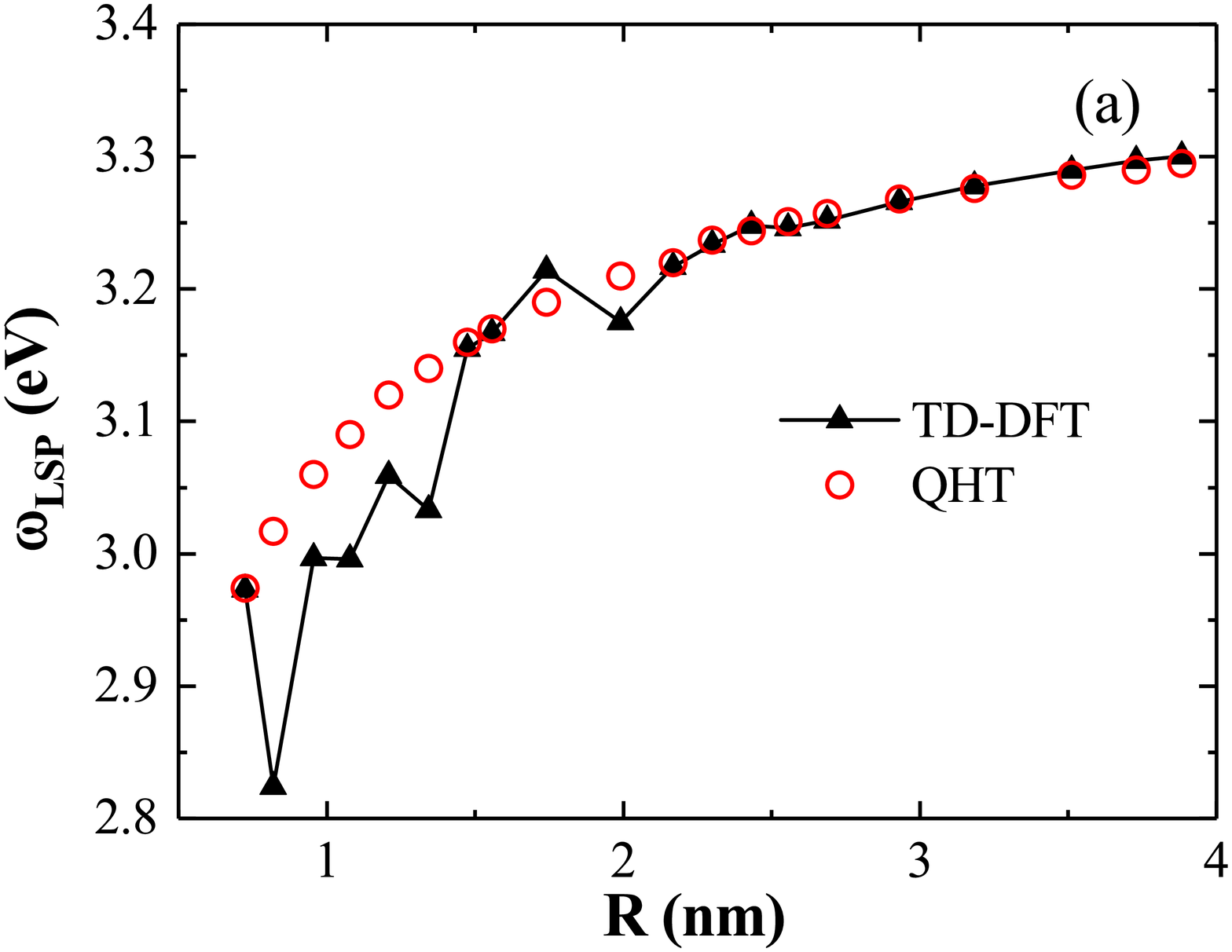}
	\includegraphics[width=8cm]{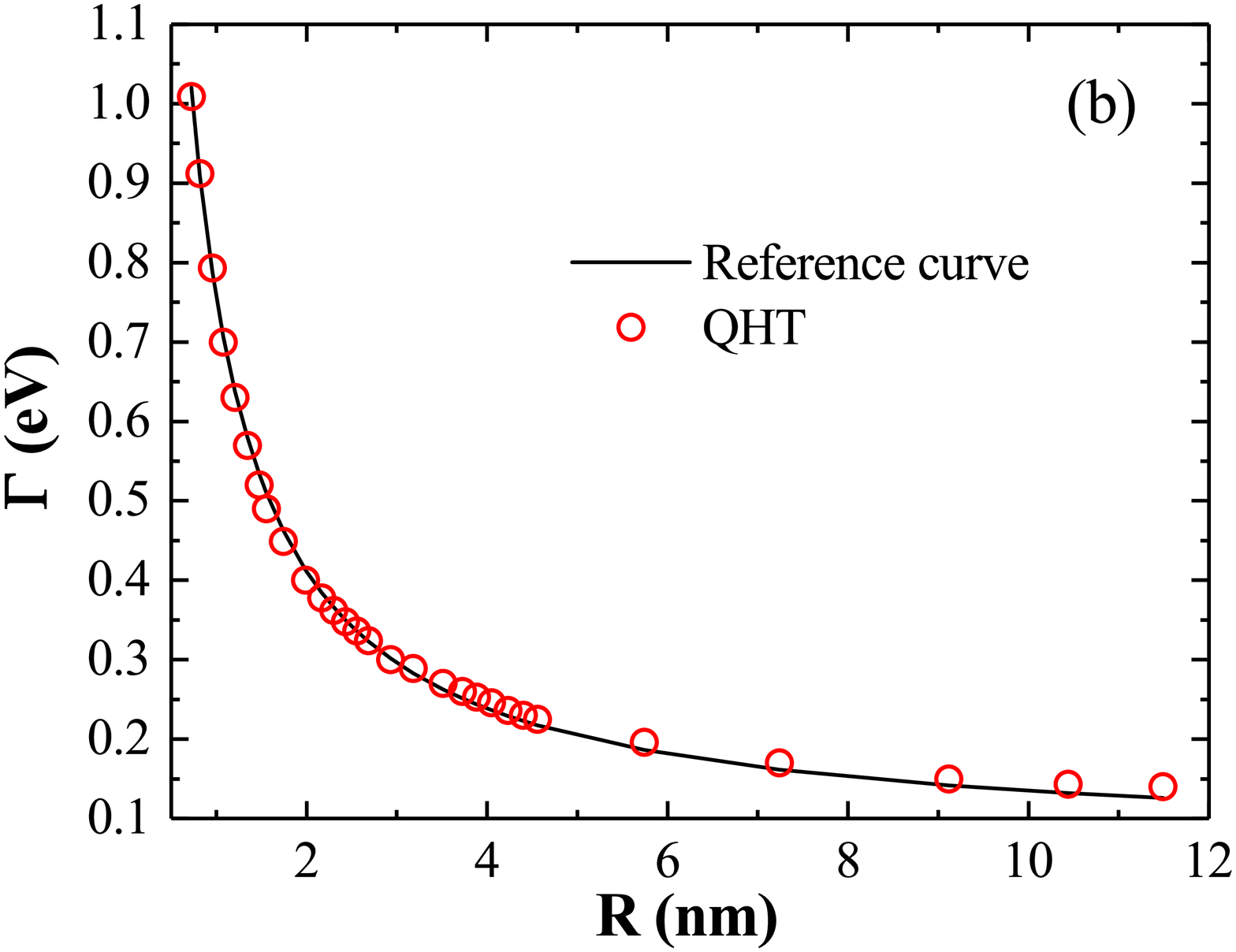}	
	\caption{(a) Resonance energy $\omega_{LSP}$, and (b) width $\Gamma$ by using the coefficient $A$ described Eq. \eqref{fitA}. For both $\omega_{LSP}$ and $\Gamma$, the agreement between our QHT results and reference values is very good. TD-DFT data are taken from Refs. \cite{PhysRevB.93.205405,PhysRevX.11.011049}. Note that the radius range in (b) is much wider than that used for obtaining the fitting function Eq. \eqref{fitA} [see Fig. \ref{fig9}]. }
	\label{fig10}%
\end{figure}

By using this coefficient $A$ [Eq. \eqref{fitA}], we report the main LSP resonance energy $\omega_{LSP}$ and the width $\Gamma$ in Fig. \ref{fig10} for nanosphere with radius over a much wide range. For the $\omega_{LSP}$ [see Fig. \ref{fig10}(a)], good agreement with the TD-DFT can also be obtained, which is similar to the case shown in Fig. \ref{fig8}(a) by using constant coefficients $A$. In this case, the MAE with respect to reference TD-DFT is very low. The third row in Table \ref{table1} [method with A from Eq. \eqref{fitA}] shows that the MAE is about $7.2\,meV$ when $N_e \geq 338$, and about $2.9\,meV$ when $N_e \geq 1074$, which are the smallest in their columns, and therefore the present method gives the best prediction of $\omega_{LSP}$. Also striking is the excellent agreement with the reference Kreibig value [see Fig. \ref{fig10}(b)]. The MAE for $\Gamma$ is about $7.6\,meV$ over a wide range of radius when $0.72\,nm\leq R \leq 4.56\,nm$ ($40\leq N_e \leq 10^4$), while it is larger than $50.0\,meV$ when taking a constant coefficient $A=0.4$ or $A=0.5$. For even larger radius range, i.e. $0.72\,nm\leq R \leq11.49\,nm$ ($40\leq N_e \leq 1.6\times 10^5$), the MAE remains very low (within $8.1\,meV$). It should be noted that Eq. \eqref{fitA} is obtained by fitting the results for $0.72nm \le R \le 1.56nm$, while the range for excellent agreement is much larger, i.e. see Fig. \ref{fig10}(b) $0.72\,nm\leq R \leq 11.49\,nm$. These results show that the present QHT provides a great degree of predictability.

\section{Application to Nanorod }\label{section3}

From the above results, we can conclude that our parameter-free QHT for plasmonics [coupled Eqs. \eqref{subeq:1} and \eqref{PEequation}] can give both the accurate main LSP resonance energy $\omega_{LSP}$ and the linewidth broadening for sodium nanosphere of various radii. The convergence problem can be solved by using the density-dependent damping rate as defined by Eq. \eqref{gaman0} with $r_q\approx10$. The vW coefficient should be ${{\lambda }_{w}}=0.4$. The diffusing $D$ is given by Eq. \eqref{densityDifu} with the coefficient $A$ described by Eq. \eqref{fitA}. 

\begin{figure}[htbp]
	\centering
	\includegraphics[width=8cm]{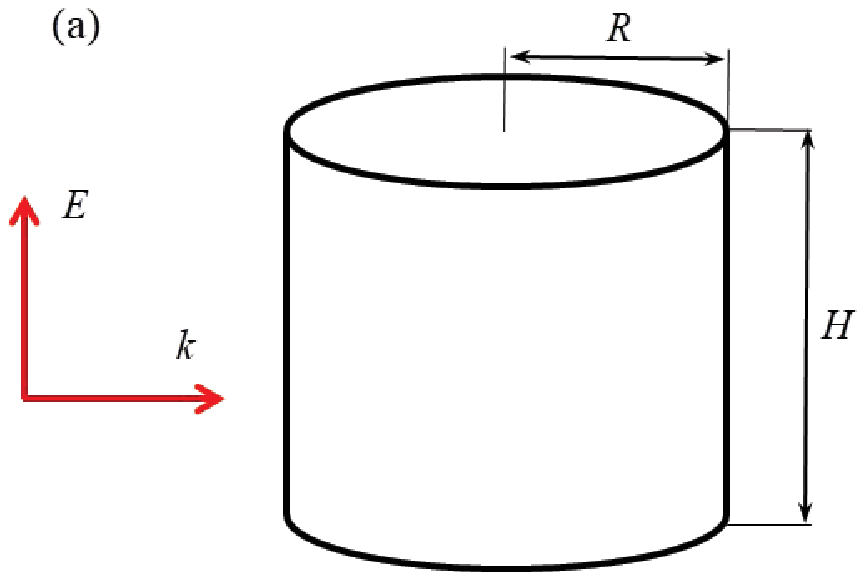}
	\includegraphics[width=8cm]{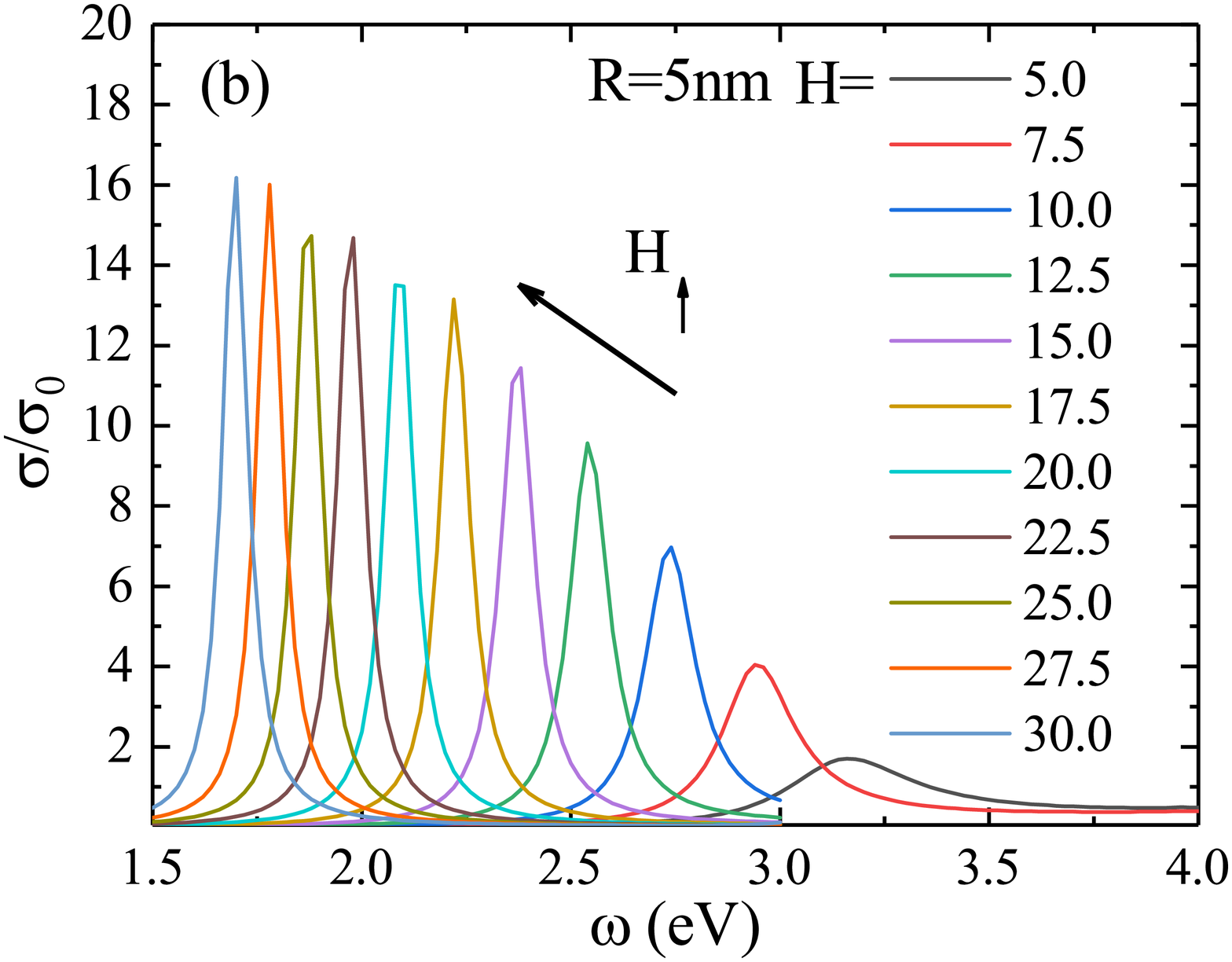}
	\caption{(a) A schematic diagram of a sodium nanorod of radius R and height H irradiated by a plane light wave polarized along the rod axis. (b) Normalized absorption cross section ($\sigma /\sigma_0$) for a jellium nanorod with radius $ R=5.0\,nm $ and the height $H$ ranging from $5.0\,nm$ to $30.0\,nm$. Here, $\sigma_0 = 2RH$ is the geometrical area. Both the resonance energy and the width decrease with increasing $H$.}
	\label{Cylindrical1}%
\end{figure}

In this section, we first apply the above QHT to investigate the optical response of sodium jellium nanorods. Then, we will show that both the resonance energy and the width are robust if the coefficient $A$ is described by Eq. \eqref{fitAR} with $R$ interpreted as several different `effective length' $R_{eff}$ of nanorod \cite{doi:10.1063/1.445794}.  

The absorption cross section is calculated for three sets of nanorods of different sizes, which are irradiated by  plane light waves polarized along the rod axis in order to effectively induce the longitudinal LSP resonance [Fig. \ref{Cylindrical1}(a)]. In the first set, the electron number $N_e$ is fixed at $186$ with the height $H$ varying from $2.0$ to $4.5\, nm$. In the second set, the radius $R=5.0\,nm$ with $H$ taking the values between $5.0$ and $30.0\,nm$. R is increased to $10.0\,nm$ in the third set and $H$ falls between $10.0$ and $60.0\,nm$. For all the three sets, the aspect ratios ($H/R$) are between $1.0$ and $7.0$. 

Figure \ref{Cylindrical1} (b) shows the normalized absorption cross section $\sigma /\sigma_0$ for nanorods with $R=5.0\,nm$. With the height $H$ increasing from $5.0\,nm$ to $30.0\,nm$, the LSP resonance energy decreases from $3.175\,eV$ to $1.697\,eV$. In addition, the spectra width decreases quickly, i.e. from $0.336\,eV$ to $0.071\,eV$. These findings are consistent with those obtained by time-dependent orbital-free density functional theory \cite{doi:10.1021/acs.jpcc.9b10510}, where the longer nanorod with a constant radius gives larger resonance wavelength and smaller linewidth. 

\begin{figure}[htbp]
	\centering
	\includegraphics[width=8cm]{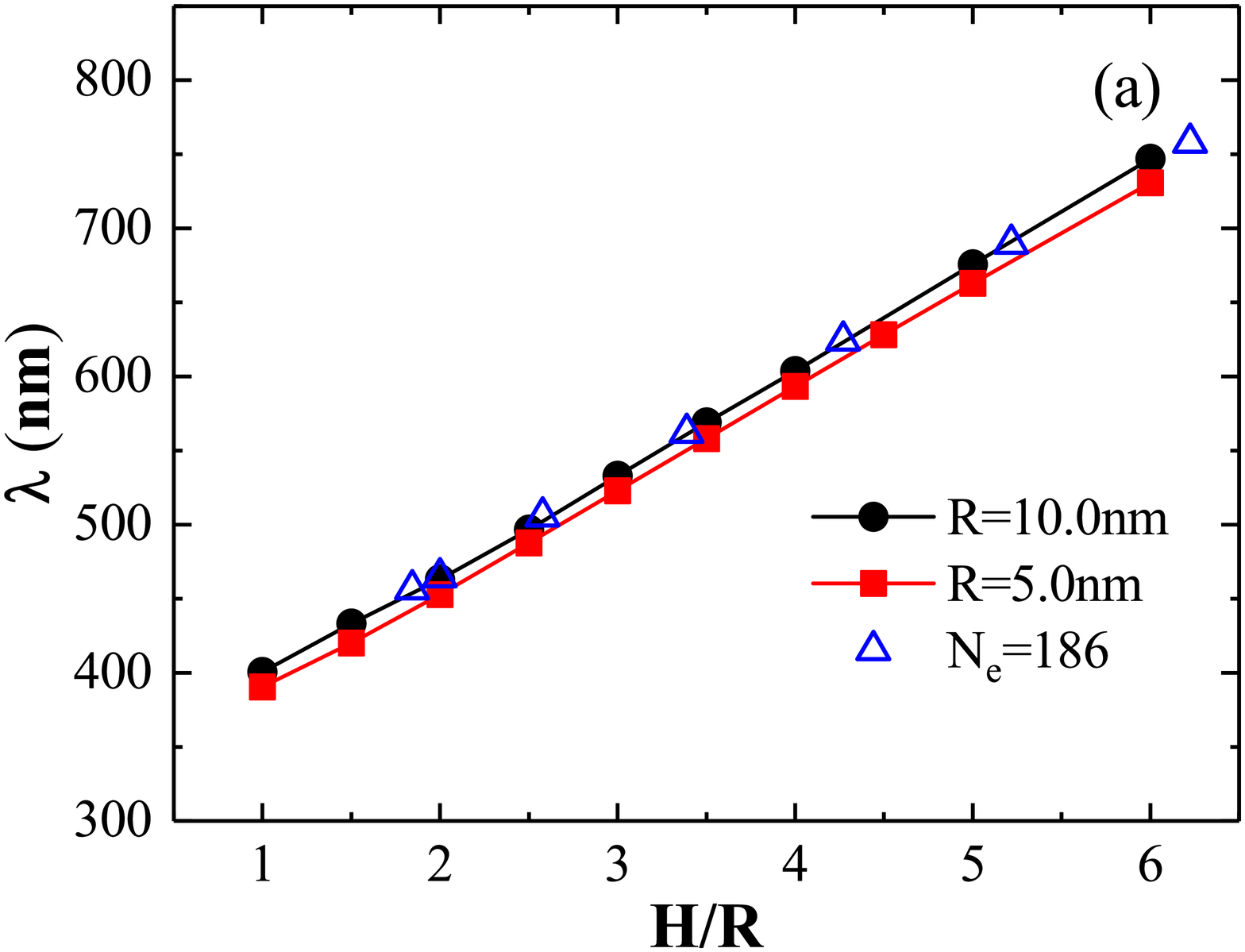}
	\includegraphics[width=8cm]{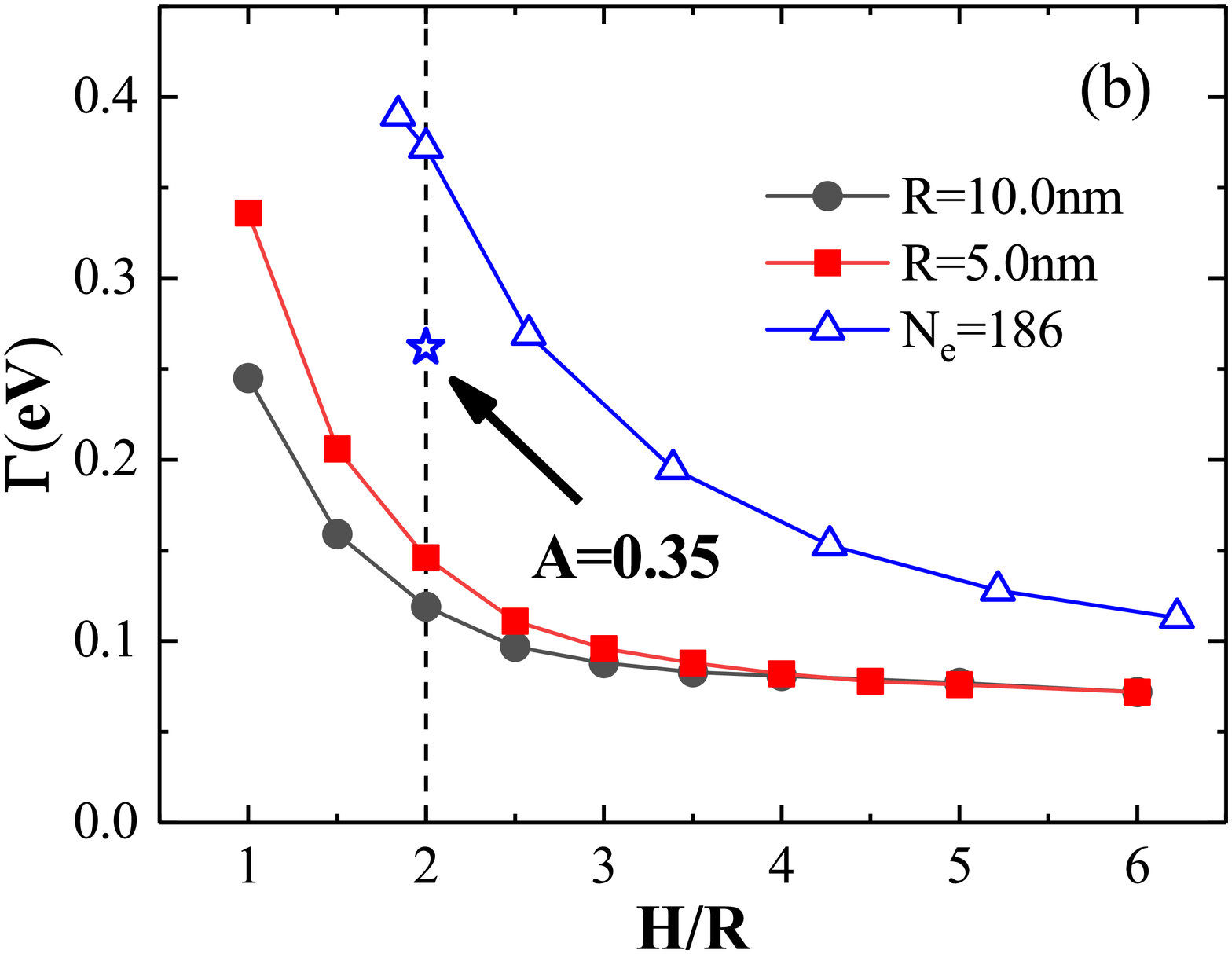}
	\caption{(a) The longitudinal LSP resonance wavelength $\lambda$, and (b) the width $\Gamma$ as a function of the aspect ratio $H/R$. Three different sets of nanorod are considered, i.e. small nanorods with constant electron number $N_e=186$ (blue triangle), nanorods with a small constant radius $R=5.0\,nm$ (red square), and relatively large constant radius $R=10.0\,nm$ (black dot). The blue star on the dashed vertical line in (b) at $H/R=2.0$ represents the the width for nanorod with $Ne=186$ when the coefficient $A$ changes to the same value for $R=5.0\,nm$ at $H/R=2.0$, i.e. $A=0.35$. }
	\label{fig12}%
\end{figure}

In Figure \ref{fig12}, we plot the longitudinal LSP resonance wavelength $\lambda$ and the width $\Gamma$ as a function of the aspect ratio. For the resonance wavelength [see Fig. \ref{fig12}(a)], it is found that all the results are located almost on the same line. Although the radius $R$ and height $H$ of the nanorods vary a lot, the longitudinal LSP resonance wavelength is determined only by the aspect ratio $H/R$ rather than $R$ and $H$ themselves. The fitting gives a linear relation $\lambda=324.55+68.61\,H/R$, which is similar to that obtained in Ref. \cite{Wen_2021} for Ag under the LRA ($\lambda=287.92+77.20\,H/R$). Intuitively, the nanorod can be regarded as quasi-one-dimensional Fabry–P\'{e}rot resonators with the charge oscillation being parallel to the rod axis for the longitudinal LSP resonance. The dipolar  resonance condition can be written as  $\lambda/2n_{eff}(\lambda,R)=H(\lambda)+2\delta(\lambda)$, in which $\delta$ is the decay length of the displacement current in vacuum, and $n_{eff}$ is the real part of the effective index for the waveguide mode. For sodium nanowire with $R=10.0\,nm$, we find a perfect linear relation $\lambda/(2n_{eff}R)=-4.35831+0.01715\lambda$ when $350.0\,nm \leq \lambda \leq 1000.0\,nm$ by using the finite-element solver COMSOL Multiphysics from a mode analysis calculation. So, a linear relation between resonance wavelength and aspect ratio can be obtained $\lambda=254.13+58.31\,H/R$, if $2\delta$ is neglected. However, the intercept $254.13$ is smaller than $324.55$ [fitted value from Fig. \ref{fig12} (a)]. The field outside becomes tightly localized on a scale being proportional to $R$, leading to $2\delta \propto R$ \cite{PhysRevB.76.035420,Wen_2021}, which can be comparable with the nanorod length. For $R=10.0\,nm$ as an example, when $H=10.0$, $15.0$, $20.0$, $25.0$, $30.0$, $35.0$, and $40.0\,nm$, it is found that $2\delta=14.72$, $15.52$, $16.25$, $17.06$, $17.95$, $18.83$, and $19.68\,nm$, respectively. Although $2\delta$ can not be neglected, the fitting gives a linear relation $2\delta=13.001+0.166H$, in which the first term can be written as $1.3001R$, and is much larger than the second term.  Combined with $\lambda/(2n_{eff}R)=-4.35831+0.01715\lambda$, one can obtain the linear relation between the resonance wavelength and the aspect ratio from the resonance condition.

Figure \ref{fig12}(b) shows the width $\Gamma$,  which decreases quickly for smaller $H/R$ and slowly for larger $H/R$. For $R=5.0\,nm$ as an example, $\Gamma$ decreases from $0.336\,eV$ to $0.096\,eV$, when $H/R$ increase from $1.0$ to $3.0$. A total decrease about $0.240\,eV$ is found. It further decreases to $0.072\,eV$ when $H/R$ increases to $6.0$. There is only a small decrease ($0.024\,eV$), which is about one tenth of the former. Since the width $0.072\,eV$ at $H/R=6.0$ is much close to the bulk value $0.066\,eV$,  it will decrease more slowly for even larger nanorod, i.e. $H/R>6.0$. Note that the aspect ratios $H/R$ are proportional to the height $H$ for all the three sets, i.e.  $H/R\propto H^{3/2}$ for $N_e=186$, and $H/R\propto H$ for both $R=5.0\,nm$ and $R=10.0\,nm$. Thus, $\Gamma$ decreases with increasing aspect ratio $H/R$ and more so for shorter nanorod. In addition, compared with the results for $R=10.0\,nm$, $\Gamma$ for nanorod with a smaller radius $R=5.0\,nm$ decrease more quickly.  When $H/R=1.0$, $\Gamma$ for $R=5.0\,nm$ is much larger than that for $R=10.0\,nm$, and they are nearly the same for $R=5.0$ and $10.0\,nm$ when $H/R>4.0$.  These results clearly show that $\Gamma$ decreases with increasing aspect ratio $H/R$ and more quickly for smaller aspect ratio, especially for smaller nanorod.

For small nanorods with the same aspect ratio, we find that the width $\Gamma$ is larger for smaller nanorod. For example, see the vertical line at $H/R=2$, we have $\Gamma=0.372\,eV$ (for $Ne=186$), which is much larger than $0.146\,eV$ (for $R=5\,nm$) and $0.119\,eV$ (for $R=10\,nm$). Such a large difference is due to the different diffusing coefficients $A$. According to Eq. \eqref{fitA}, we have $A=0.48$ for $Ne=186$, $A=0.35$ for $R=5\,nm$, and $A=0.33$ for $R=10\,nm$, where the first one is much larger than the last two. If the coefficient $A$ for the nanorod with $Ne=186$ decrease from $A=0.48$ to $0.35$ (the same value for $R=5\,nm$ at $H/R=2$), the width can be drastically reduced from $0.372\,eV$ to $0.262\,eV$ [see the star on the vertical line \ref{fig12}(b)]. However, in this case, $\Gamma$ remains larger than that for $R=5\,nm$. Similar different width between $R=5\,nm$ and $R=10\,nm$ can be seen, although their coefficients $A$ are nearly the same. Thus, the nanorod height (radius) has great influence on the width $\Gamma$.

From the above results, we find that the width $\Gamma$ decrease with increasing aspect ratio $H/R$. Under the same aspect ratio, it decreases with increasing height $H$. Usually, the size dependent broadening can be described by $\Gamma=\gamma_0 +A_1v_F/L_{eff}$ with $L_{eff}$ being the effective confinement length \cite{doi:10.1063/1.445794,doi:10.1021/nl400777y,B604856K,C4CS00131A,doi:10.1021/jp9917648}. The interpretation of $L_{eff}$ differs slightly. It is approximated by $0.65V/S$ with $V$ and $S$ being the volume of the particle and the projected area perpendicular to the direction of the applied field, respectively, which leads to $L_{eff}\propto H$ for the longitudinal LSP mode of nanorod \cite{doi:10.1063/1.445794}. For individual gold nanorods protected by a silica shell, it was found that $L_{eff}=H(H/R)^\beta$ with $\beta=-0.5$ reproducing the measured results \cite{doi:10.1021/nl400777y}. However, for both cases, the effective confinement length $L_{eff}$ increases with increasing height $H$, which is consistent with our observation that the width $\Gamma$ decreases with increasing $H$. For sodium nanorod, a rational form for $L_{eff}$ should be $L_{eff}=H(H/R)^\beta$, which is similar to Ref. \cite{doi:10.1021/nl400777y} where $\beta=-0.5$ for silica-coated gold nanorod. By fitting the three sets of results with $\Gamma=\gamma_0 +A_1v_F/(H(H/R)^\beta)$ , we have $A_1=1.62$ and $\beta =0.88$ for $Ne=186$, $A_1=1.99$ and $\beta =0.88$ for $R=5.0\,nm$, and $A_1=2.62$ and $\beta =0.80$ for $R=10.0\,nm$. Different from Ref. \cite{doi:10.1021/nl400777y} where $\beta=-0.5$ for silica-coated gold nanorod, we have a positive $\beta$ and they are all around $0.8$ for the three sets of sodium nanorods. These results clearly show that $\Gamma$ decreases with increasing aspect ratio $H/R$ and more quickly for smaller $H/R$. Under the same aspect ratio, it decreases with increasing height $H$.

The above results are obtained with $A$ defined through the total electron number $N_e$ [Eq. \eqref{fitA}], which is irrelevant to the aspect ratio of the nanorod. In the following, we will discussion the effect of the coefficient $A$ described by Eq. \eqref{fitAR} based on  different interpretations of the radius $R$.  For nanosphere, the same $A$ can be obtained by using either the radius $R$ [Eq. \eqref{fitAR}] or the total electron $N_e$ [Eq. \eqref{fitA}]. But for a nanorod, it is hard to define $R$. For nanosphere, the effective confinement length can be written as $L_{eff}=0.86R$, which leads to $R=L_{eff}/0.86$ \cite{doi:10.1063/1.445794}. Thus, the radius $R$ in Eq. \eqref{fitAR} can be replaced by an effective radius $R_{eff}=L_{eff}/0.86$. If this relation is also true for nanorod, one can obtain an effective radius
\begin{equation}\label{Reff2}
R_{eff}=1.36H,
\end{equation}
since the effective confinement length for the longitudinal LSP resonance of a nanorod is $L_{eff}=0.631H$ \cite{doi:10.1063/1.445794}.

Another possible interpretation for $R$ is based on purely classical descriptions of surface scattering, where $R$ for a nanosphere is the average chord length between one point on the surface and any other point, averaged over all points on the surface as the initial point [see Eq. (1.1) in Ref. \cite{doi:10.1063/1.445794}]. For the nanorod, this average length is the height $H$, since the electron moves along the rod axis for the longitudinal LSP resonance and the collisions take place on the two bottom surfaces. It is $V/S=H$ [see Eq. (4.1) in Ref. \cite{doi:10.1063/1.445794}]. In this case, the effective radius becomes 
\begin{equation}\label{Reff3}
R_{eff}=H.
\end{equation}

\begin{figure}[htbp]
	\centering
	\includegraphics[width=8cm]{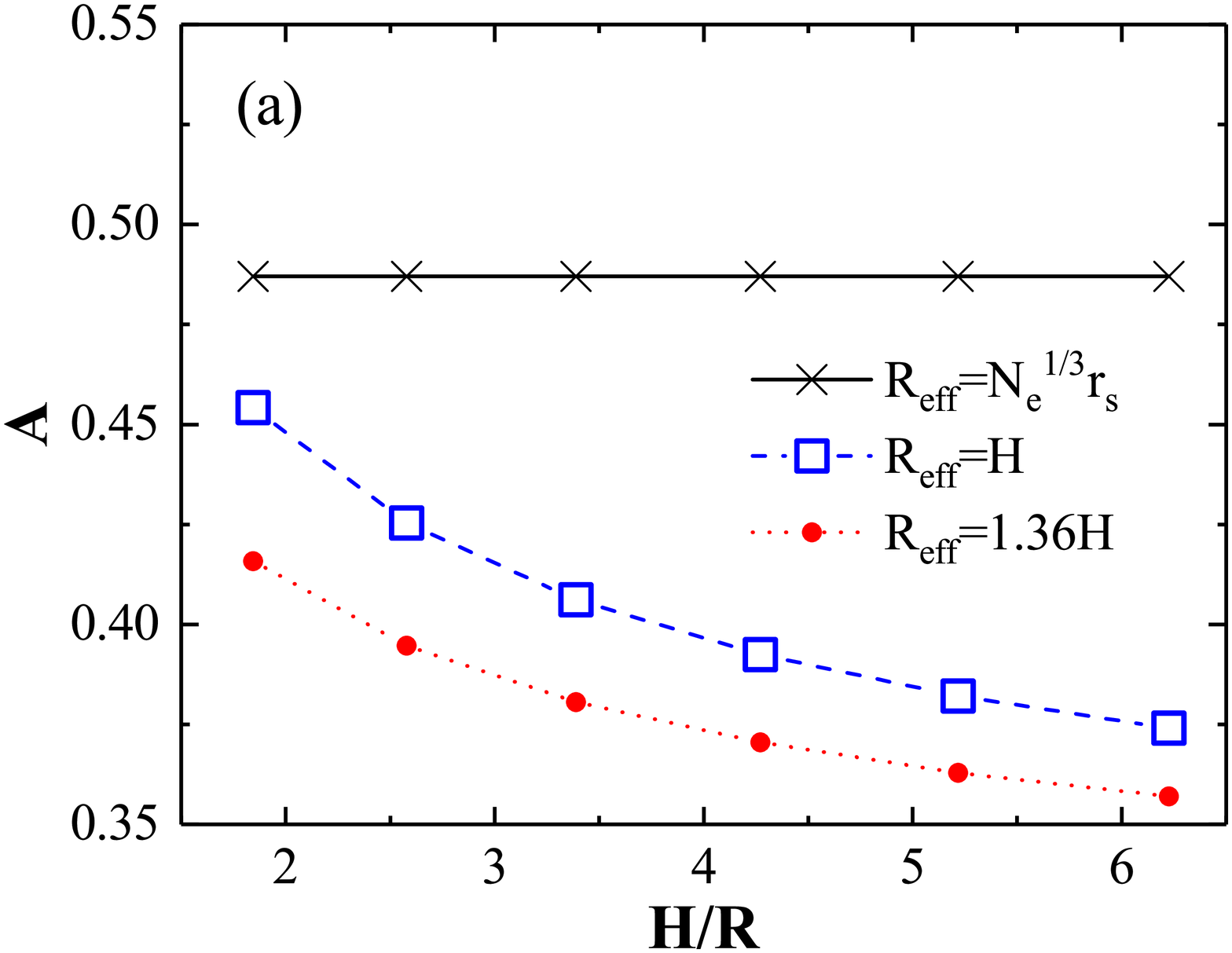}
	\includegraphics[width=8cm]{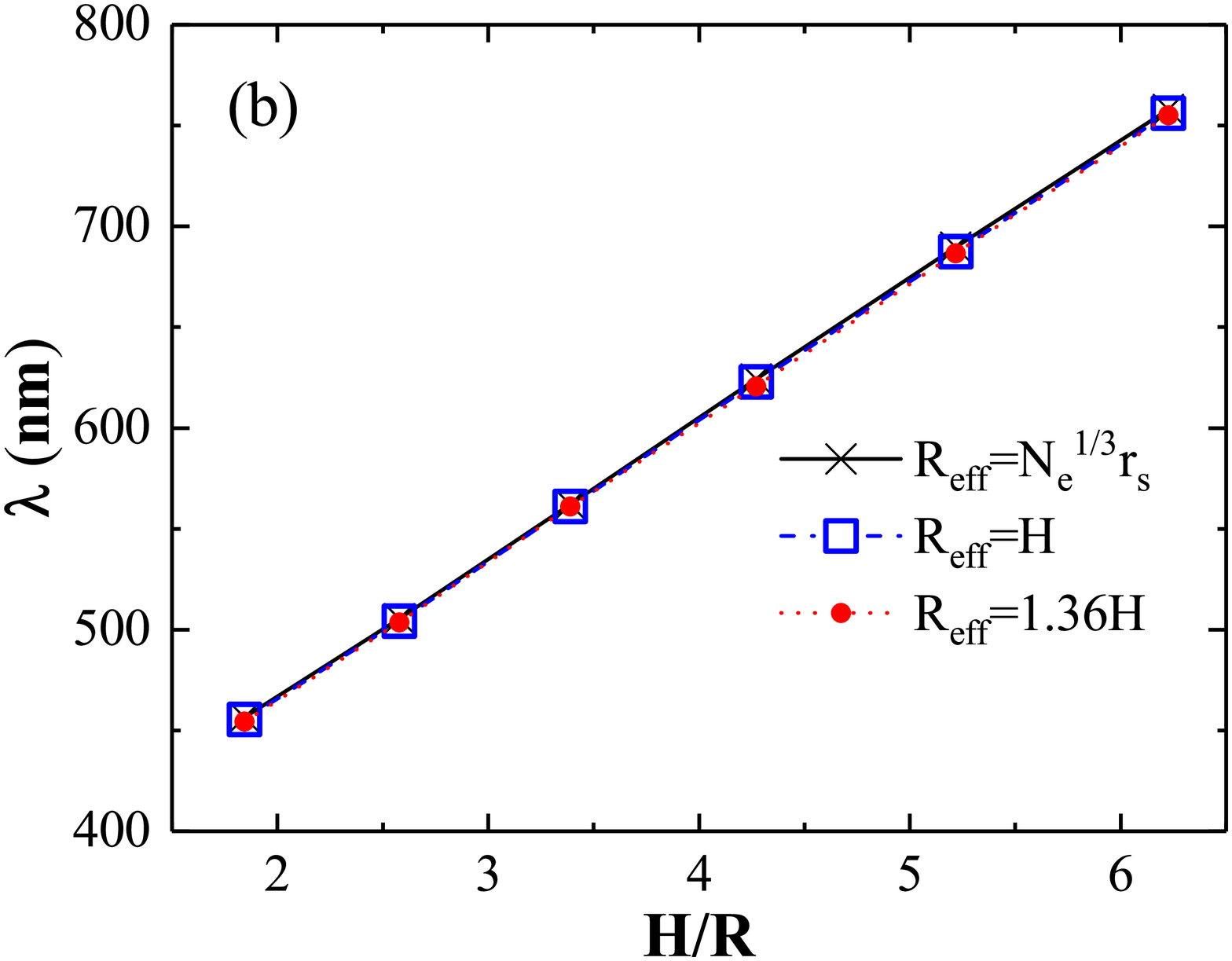}
	\includegraphics[width=8cm]{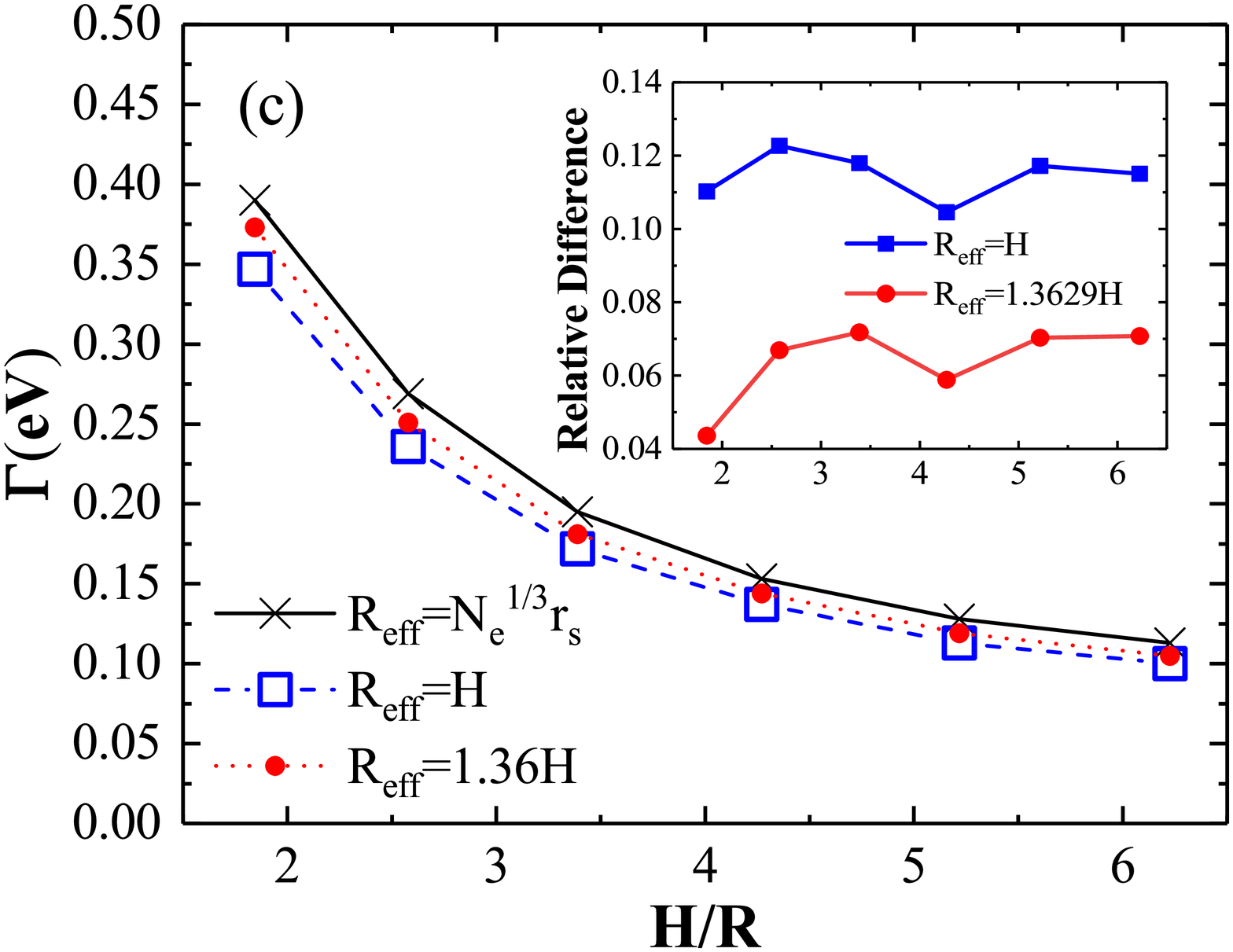}
	\caption{(a) Coefficient $A$, (b) resonance wavelength $\lambda$, and (c) width $\Gamma$ by using three different effective radii. Robustness and good predictability for our QHT are demonstrated if the coefficient $A$ in Eq. \eqref{fitA} is described by the effective radius (see in the text) for nanorod, i.e.  $R_{eff}=N_{e}^{1/3}r_s$ (black crosses on solid line), $R_{eff}=H$ (blue squares on dashed line), and  $R_{eff}=1.36H$ (red circles on dotted line). Here, the sets with $N_e=186$ is used, which shows the largest difference among the three sets.  The insets in (c) shows the relative difference for $ R_{eff}=H$ (blue square) and $R_{eff}=1.36H $ (red circle) taking the width obtained by using  $ R_{eff}=N_{e}^{1/3}r_s$ as a reference. }
	\label{fig13}%
\end{figure}

According to the above discussion,  there may be three different effective radius for a nanorod, i.e. $R_{eff}=N_e^{1/3}r_s$, which is the same as $A$ defined by $N_e$ (the equivalent radius for a nanosphere with the same volume of the nanorod), $R_{eff}=1.36H$ [Eq. \eqref{Reff2}]), and $R_{eff}=H$ [Eq. \eqref{Reff3}]. By taking $R_{eff}$ in place of $R$ in Eq. \eqref{fitAR}, i.e. $A=0.31+3.94a_0/R_{eff}$, we see that the coefficient $A$ changes little due to the different choice of $R_{eff}$ when $R=5.0\,nm$ (within $0.03$) and $R=10.0\,nm$ (within $0.02$). The resonance wavelength and the width will not be affected. But for the set with $N_e=186$, the size dependent part $3.94a_0/R_{eff}$ has relatively large effect. Below, we will focus on this set ($N_e=186$). Figure \ref{fig13}(a) shows the coefficients $A$ as a function of the aspect ratio $H/R$ based on the three definitions of $R_{eff}$. Instead of being a constant value $A=0.487$ for $R_{eff}=N_e^{1/3}r_s$, it changes from $0.454$ to $0.374$ for $R_{eff}=H$, and from $0.416$ to $0.357$ for $R_{eff}=1.36H$. The maximum difference among them is about $0.130$. 

The resonance wavelength and the width as a function of the aspect ratio are calculated using the three effective radii $R_{eff}$, as shown in Figs. \ref{fig13}(b) and (c). It can be seen that the resonance wavelength are the same for all three cases [see Fig. \ref{fig13}(b)], confirming that the coefficient $A$ has less influence on the resonance energy. For the width [see Fig. \ref{fig13}(c)], it is also nearly independent on the definition of $R_{eff}$. The width predicted using $R_{eff}=N_e^{1/3}r_s$ is a little larger than that using $R_{eff}=1.36H$ and $R_{eff}=H$ by $6\%$ and $11\%$, respetively [see the inset in Fig. \ref{fig13}(c)].  These results clearly show that our method is robust against to $R_{eff}$ and offers a great degree of predictability.

\section{CONCLUSION}
In conclusion, we have formulated the self-consistent quantum hydrodynamic theory as a parameter-free form [coupled Eqs. \eqref{subeq:1} and \eqref{PEequation}], by which both the resonance energy and linewidth can be obtained for nanostructure of arbitrary shape. Quantum effects such as electronic spill-out and Landau damping are taken into account. Both the ground and excited states have been solved by using the same energy functional, where the KE is described by $\mathrm{TF}\lambda_w \mathrm{vW}$. We have found that the fraction of the vW potential $\lambda_w$ has great effect on the resonance energy and the diffusing has been quite successful in describing the size dependent broadening. For sodium jellium nansosphere, there are three main findings. 

Firstly, the damping given by Eq. \eqref{gaman0} with $r_q\approx10$  can be used to solve the convergence problem. Inside and around the nanospere, the damping is the same as the bulk term $\gamma_0$, which has no influence on the main LSP resonance. But in the density-tail region, it increases exponentially with position away from the metal suface $\gamma \left( r \right)\propto n_0^{-5/6}$, which damps the delocalized state efficiently, i.e. the computation-size dependent spurious peaks at energies higher than the main LSP resonance. 

Secondly, we have numerically determined that the fraction of the vW potential should be $\lambda_w\approx0.4$ in order to give the correct main LSP resonance energy. It is well known that $\lambda_w$ controls the degree of the electron spill-out, with smaller $\lambda_w$ corresponding to a less spill-out. However, its exact value is not well defined and usually in the range of $1/9$ to $1$. For nanosphere with $N_e=438$, we have numerically proved that the resonance energy of the main LSP varies linearly with $\lambda_w$. The slope is very large with a value around $0.4\,eV$. Thus, the fraction of the vW potential should be around 0.4 in order to give the TD-DFT resonance energy. It should be noted that our value is pretty around $0.435$, which has been used to give the work function close to the DFT value. By using  $\lambda_w=0.4$ for nanospheres with various radii, we have found that it offers a great degree of predictability. It gives rise to the same LSP resonance energy as with the TD-DFT, with the results by the present QHT marking almost exactly the mean trajectory of TD-DFT data. When $R\geq 2.16\,nm$ ($N_e\geq1074$ ), the mean average errors with respect to reference TD-DFT is about $2.9\,meV$. 

Lastly, we have shown that the size dependent broadening can be treated properly by using the diffusing $D$ described in Eq. \eqref{densityDifu} with the coefficient $A$ given by Eq. \eqref{fitA}. It is noteworthy that although the coefficient $A$ [Eq. \eqref{fitA}] is obtained by fitting the results for small nanosphere radius, i.e. $0.72nm \le R \le 1.56nm$, it yields the same SP linewidth broadening as with the Kreibig approach for nanosphere with radius over a much larger range, i.e. see Fig. \ref{fig10}(b) $ R \approx 12nm$. Thus, our method by using the diffusing $D$ [Eq. \eqref{densityDifu}] with the coefficient $A$ given by Eq. \eqref{fitA} offers a great degree of predictability. 

By applying our parameter-free QHT to sodium jellium nanorods, we have found that there exists a perfect linear relation between the main longitudinal LSP resonance wavelength and the aspect radio for nanorods of various sizes, i.e. $\lambda=324.55+68.61\,H/R$. The size dependent broadening can be described well by $\Gamma=\gamma_0+av_F/L_{eff}$. Here, $L_{eff}=H(H/R)^\beta$ with $\beta$ around $0.8$. Thus, the width decreases with increasing aspect ratio $H/R$ and height $H$. We have also shown that the coefficient $A$ given by Eq. \eqref{fitA} through the total electron number is robust and provides a great degree of predictability.

We believe this work offers a valid and efficient solution for studying metal nanostructure of relatively large size and simultaneously of arbitrary shape. Optical response of nanoparticle dimers, disks, film-coupled nanoparticles or macroscopic systems can be efficiently obtained. Meanwhile, quantum light-matter interactions, for example see Refs. \cite{RN989,https://doi.org/10.1002/lpor.201700113,Wen:20,PhysRevA.99.053844,PhysRevA.85.053827,lu2021plasmonic,doi:10.1063/5.0033531,Karanikolas:21,thanopulos2020non,moazzezi2020second,hamza2021frster,Lassalle:18}, have been the subject of intense theoretical and practical interest. Our method can serve as a robust and valuable tool in this field, when plasmonic effects in mesoscopic metal nanostructure is exploited.

\begin{acknowledgments}
This work was supported by the National Natural Science Foundation of China (Grant Nos. 11964010, 11874315, 11464013, and 11464014), the Natural Science Foundation of Hunan Province (Grant No. 2020JJ4495), the Scientific Research Fund of Hunan Provincial Education Department (Grant No. 21A0333), and Hunan Provincial Innovation Foundation For Postgraduate (Grant Nos. CX20211038, and CX2018B706).
\end{acknowledgments}

\bibliography{refer11}

\end{document}